\documentclass[twocolumn]{aastex701}

\usepackage{mathtools}
\usepackage{xspace}
\usepackage{dcolumn}
\usepackage{subcaption}

\newcommand{\Ntot}{10,262\xspace}

\newcommand{\Nbins}{14\xspace}
\newcommand{\NbinGals}{4050\xspace}

\newcommand{\TFslope}{$-7.22\pm0.01$~AB~mag\xspace}
\newcommand{\TFscatter}{$0.466\pm0.001$~AB~mag\xspace}

\shorttitle{DESI DR1 Tully-Fisher Catalog}
\shortauthors{Douglass et al.}

\begin{document}

\title{The DESI DR1 Peculiar Velocity Survey: The Tully-Fisher Distance Catalog}

\correspondingauthor{K.~Douglass}
\email{kellyadouglass@rochester.edu}


\author[0000-0002-9540-546X]{K.~Douglass}
\affiliation{Department of Physics \& Astronomy, University of Rochester, 206 Bausch and Lomb Hall, P.O. Box 270171, Rochester, NY 14627-0171, USA}
\email{kellyadouglass@rochester.edu}

\author[0000-0001-5537-4710]{S.~BenZvi}
\affiliation{Department of Physics \& Astronomy, University of Rochester, 206 Bausch and Lomb Hall, P.O. Box 270171, Rochester, NY 14627-0171, USA}
\email{sbenzvi@ur.rochester.edu}

\author[0000-0001-6315-8743]{A.~G.~Kim}
\affiliation{Lawrence Berkeley National Laboratory, 1 Cyclotron Road, Berkeley, CA 94720, USA}
\email{agkim@lbl.gov}

\author[0000-0002-1988-1747]{S.~Moore}
\affiliation{Department of Physics \& Astronomy, University of Rochester, 206 Bausch and Lomb Hall, P.O. Box 270171, Rochester, NY 14627-0171, USA}
\email{smoore55@ur.rochester.edu}

\author[0000-0003-4074-5659]{A.~Carr}
\affiliation{Korea Astronomy and Space Science Institute, 776, Daedeokdae-ro, Yuseong-gu, Daejeon 34055, Republic of Korea}
\email{anthonycarr@kasi.re.kr}

\author[0000-0001-5532-561X]{J.~Largett}
\affiliation{Department of Physics \& Astronomy, University of Rochester, 206 Bausch and Lomb Hall, P.O. Box 270171, Rochester, NY 14627-0171, USA}
\email{jlargett@u.rochester.edu}

\author[0000-0002-4248-2840]{N.~Ravi}
\affiliation{Department of Physics \& Astronomy, University of Rochester, 206 Bausch and Lomb Hall, P.O. Box 270171, Rochester, NY 14627-0171, USA}
\email{nravi3@ur.rochester.edu}

\author{J.~Aguilar}
\affiliation{Lawrence Berkeley National Laboratory, 1 Cyclotron Road, Berkeley, CA 94720, USA}
\email{jaguilar@lbl.gov}

\author[0000-0001-6098-7247]{S.~Ahlen}
\affiliation{Department of Physics, Boston University, 590 Commonwealth Avenue, Boston, MA 02215 USA}
\email{ahlen@bu.edu}

\author[0000-0003-3433-2698]{A.~J.~Amsellem}
\affiliation{Department of Physics, Carnegie Mellon University, 5000 Forbes Avenue, Pittsburgh, PA 15213, USA}
\email{aamselle@andrew.cmu.edu}

\author{J.~Bautista}
\affiliation{Aix Marseille Univ, CNRS/IN2P3, CPPM, Marseille, France}
\email{bautista@cppm.in2p3.fr}

\author[0000-0001-9712-0006]{D.~Bianchi}
\affiliation{Dipartimento di Fisica ``Aldo Pontremoli'', Universit\`a degli Studi di Milano, Via Celoria 16, I-20133 Milano, Italy}
\affiliation{INAF-Osservatorio Astronomico di Brera, Via Brera 28, 20122 Milano, Italy}
\email{davide.bianchi1@unimi.it}

\author[0000-0002-5423-5919]{C.~Blake}
\affiliation{Centre for Astrophysics \& Supercomputing, Swinburne University of Technology, P.O. Box 218, Hawthorn, VIC 3122, Australia}
\email{cblake@astro.swin.edu.au}

\author{D.~Brooks}
\affiliation{Department of Physics \& Astronomy, University College London, Gower Street, London, WC1E 6BT, UK}
\email{david.brooks@ucl.ac.uk}

\author{T.~Claybaugh}
\affiliation{Lawrence Berkeley National Laboratory, 1 Cyclotron Road, Berkeley, CA 94720, USA}
\email{tmclaybaugh@lbl.gov}

\author[0000-0002-2169-0595]{A.~Cuceu}
\affiliation{Lawrence Berkeley National Laboratory, 1 Cyclotron Road, Berkeley, CA 94720, USA}
\email{acuceu@lbl.gov}

\author[0000-0002-1769-1640]{A.~de la Macorra}
\affiliation{Instituto de F\'{\i}sica, Universidad Nacional Aut\'{o}noma de M\'{e}xico,  Circuito de la Investigaci\'{o}n Cient\'{\i}fica, Ciudad Universitaria, Cd. de M\'{e}xico  C.~P.~04510,  M\'{e}xico}
\email{macorra@fisica.unam.mx}

\author[0000-0002-7852-167X]{R.~Demina}
\affiliation{Department of Physics \& Astronomy, University of Rochester, 206 Bausch and Lomb Hall, P.O. Box 270171, Rochester, NY 14627-0171, USA}
\email{regina@pas.rochester.edu}

\author{P.~Doel}
\affiliation{Department of Physics \& Astronomy, University College London, Gower Street, London, WC1E 6BT, UK}
\email{apd@star.ucl.ac.uk}

\author[0000-0003-4992-7854]{S.~Ferraro}
\affiliation{Lawrence Berkeley National Laboratory, 1 Cyclotron Road, Berkeley, CA 94720, USA}
\affiliation{University of California, Berkeley, 110 Sproul Hall \#5800 Berkeley, CA 94720, USA}
\email{sferraro@lbl.gov}

\author[0000-0002-3033-7312]{A.~Font-Ribera}
\affiliation{Institut de F\'{i}sica d’Altes Energies (IFAE), The Barcelona Institute of Science and Technology, Edifici Cn, Campus UAB, 08193, Bellaterra (Barcelona), Spain}
\email{afont@ifae.es}

\author[0000-0002-2890-3725]{J.~E.~Forero-Romero}
\affiliation{Departamento de F\'isica, Universidad de los Andes, Cra. 1 No. 18A-10, Edificio Ip, CP 111711, Bogot\'a, Colombia}
\affiliation{Observatorio Astron\'omico, Universidad de los Andes, Cra. 1 No. 18A-10, Edificio H, CP 111711 Bogot\'a, Colombia}
\email{je.forero@uniandes.edu.co}

\author[0000-0001-9632-0815]{E.~Gaztañaga}
\affiliation{Institut d'Estudis Espacials de Catalunya (IEEC), c/ Esteve Terradas 1, Edifici RDIT, Campus PMT-UPC, 08860 Castelldefels, Spain}
\affiliation{Institute of Cosmology and Gravitation, University of Portsmouth, Dennis Sciama Building, Portsmouth, PO1 3FX, UK}
\affiliation{Institute of Space Sciences, ICE-CSIC, Campus UAB, Carrer de Can Magrans s/n, 08913 Bellaterra, Barcelona, Spain}
\email{gaztanaga@gmail.com}

\author[0000-0003-3142-233X]{S.~Gontcho A Gontcho}
\affiliation{Lawrence Berkeley National Laboratory, 1 Cyclotron Road, Berkeley, CA 94720, USA}
\affiliation{University of Virginia, Department of Astronomy, Charlottesville, VA 22904, USA}
\email{satya@virginia.edu}

\author{G.~Gutierrez}
\affiliation{Fermi National Accelerator Laboratory, PO Box 500, Batavia, IL 60510, USA}
\email{gaston@fnal.gov}

\author[0000-0001-9822-6793]{J.~Guy}
\affiliation{Lawrence Berkeley National Laboratory, 1 Cyclotron Road, Berkeley, CA 94720, USA}
\email{jguy@lbl.gov}

\author[0000-0002-9136-9609]{H.~K.~Herrera-Alcantar}
\affiliation{Institut d'Astrophysique de Paris. 98 bis boulevard Arago. 75014 Paris, France}
\affiliation{IRFU, CEA, Universit\'{e} Paris-Saclay, F-91191 Gif-sur-Yvette, France}
\email{herreraa@iap.fr}

\author[0000-0002-6550-2023]{K.~Honscheid}
\affiliation{Center for Cosmology and AstroParticle Physics, The Ohio State University, 191 West Woodruff Avenue, Columbus, OH 43210, USA}
\affiliation{Department of Physics, The Ohio State University, 191 West Woodruff Avenue, Columbus, OH 43210, USA}
\affiliation{The Ohio State University, Columbus, 43210 OH, USA}
\email{kh@physics.osu.edu}

\author[0000-0002-1081-9410]{C.~Howlett}
\affiliation{School of Mathematics and Physics, University of Queensland, Brisbane, QLD 4072, Australia}
\email{c.howlett@uq.edu.au}

\author[0000-0001-6558-0112]{D.~Huterer}
\affiliation{Department of Physics, University of Michigan, 450 Church Street, Ann Arbor, MI 48109, USA}
\affiliation{University of Michigan, 500 S. State Street, Ann Arbor, MI 48109, USA}
\email{huterer@umich.edu}

\author[0000-0002-6024-466X]{M.~Ishak}
\affiliation{Department of Physics, The University of Texas at Dallas, 800 W. Campbell Rd., Richardson, TX 75080, USA}
\email{mishak@utdallas.edu}

\author[0000-0003-0201-5241]{R.~Joyce}
\affiliation{NSF NOIRLab, 950 N. Cherry Ave., Tucson, AZ 85719, USA}
\email{richard.joyce@noirlab.edu}

\author[0000-0001-6356-7424]{A.~Kremin}
\affiliation{Lawrence Berkeley National Laboratory, 1 Cyclotron Road, Berkeley, CA 94720, USA}
\email{akremin@lbl.gov}

\author{O.~Lahav}
\affiliation{Department of Physics \& Astronomy, University College London, Gower Street, London, WC1E 6BT, UK}
\email{o.lahav@ucl.ac.uk}

\author[0000-0002-6731-9329]{C.~Lamman}
\affiliation{The Ohio State University, Columbus, 43210 OH, USA}
\email{lamman.1@osu.edu}

\author[0000-0003-1838-8528]{M.~Landriau}
\affiliation{Lawrence Berkeley National Laboratory, 1 Cyclotron Road, Berkeley, CA 94720, USA}
\email{mlandriau@lbl.gov}

\author[0000-0001-7178-8868]{L.~Le~Guillou}
\affiliation{Sorbonne Universit\'{e}, CNRS/IN2P3, Laboratoire de Physique Nucl\'{e}aire et de Hautes Energies (LPNHE), FR-75005 Paris, France}
\email{llg@lpnhe.in2p3.fr}

\author[0000-0002-3677-3617]{A.~Leauthaud}
\affiliation{Department of Astronomy and Astrophysics, UCO/Lick Observatory, University of California, 1156 High Street, Santa Cruz, CA 95064, USA}
\affiliation{Department of Astronomy and Astrophysics, University of California, Santa Cruz, 1156 High Street, Santa Cruz, CA 95065, USA}
\email{alexie@ucsc.edu}

\author[0000-0003-1887-1018]{M.~E.~Levi}
\affiliation{Lawrence Berkeley National Laboratory, 1 Cyclotron Road, Berkeley, CA 94720, USA}
\email{melevi@lbl.gov}

\author[0000-0003-4962-8934]{M.~Manera}
\affiliation{Departament de F\'{i}sica, Serra H\'{u}nter, Universitat Aut\`{o}noma de Barcelona, 08193 Bellaterra (Barcelona), Spain}
\affiliation{Institut de F\'{i}sica d’Altes Energies (IFAE), The Barcelona Institute of Science and Technology, Edifici Cn, Campus UAB, 08193, Bellaterra (Barcelona), Spain}
\email{mmanera@ifae.es}

\author[0000-0002-4279-4182]{P.~Martini}
\affiliation{Center for Cosmology and AstroParticle Physics, The Ohio State University, 191 West Woodruff Avenue, Columbus, OH 43210, USA}
\affiliation{Department of Astronomy, The Ohio State University, 4055 McPherson Laboratory, 140 W 18th Avenue, Columbus, OH 43210, USA}
\affiliation{The Ohio State University, Columbus, 43210 OH, USA}
\email{martini.10@osu.edu}

\author[0000-0002-1125-7384]{A.~Meisner}
\affiliation{NSF NOIRLab, 950 N. Cherry Ave., Tucson, AZ 85719, USA}
\email{aaron.meisner@noirlab.edu}

\author{R.~Miquel}
\affiliation{Instituci\'{o} Catalana de Recerca i Estudis Avan\c{c}ats, Passeig de Llu\'{\i}s Companys, 23, 08010 Barcelona, Spain}
\affiliation{Institut de F\'{i}sica d’Altes Energies (IFAE), The Barcelona Institute of Science and Technology, Edifici Cn, Campus UAB, 08193, Bellaterra (Barcelona), Spain}
\email{rmiquel@ifae.es}

\author[0000-0002-2733-4559]{J.~Moustakas}
\affiliation{Department of Physics and Astronomy, Siena University, 515 Loudon Road, Loudonville, NY 12211, USA}
\email{jmoustakas@siena.edu}

\author{A.~Muñoz-Gutiérrez}
\affiliation{Instituto de F\'{\i}sica, Universidad Nacional Aut\'{o}noma de M\'{e}xico,  Circuito de la Investigaci\'{o}n Cient\'{\i}fica, Ciudad Universitaria, Cd. de M\'{e}xico  C.~P.~04510,  M\'{e}xico}
\email{andreamgtz@ciencias.unam.mx}

\author[0000-0001-9070-3102]{S.~Nadathur}
\affiliation{Institute of Cosmology and Gravitation, University of Portsmouth, Dennis Sciama Building, Portsmouth, PO1 3FX, UK}
\email{seshadri.nadathur@port.ac.uk}

\author[0000-0003-3188-784X]{N.~Palanque-Delabrouille}
\affiliation{IRFU, CEA, Universit\'{e} Paris-Saclay, F-91191 Gif-sur-Yvette, France}
\affiliation{Lawrence Berkeley National Laboratory, 1 Cyclotron Road, Berkeley, CA 94720, USA}
\email{npalanque-delabrouille@lbl.gov}

\author{A.~Palmese}
\affiliation{Department of Physics, Carnegie Mellon University, 5000 Forbes Avenue, Pittsburgh, PA 15213, USA}
\email{apalmese@andrew.cmu.edu}

\author[0000-0002-0644-5727]{W.~J.~Percival}
\affiliation{Department of Physics and Astronomy, University of Waterloo, 200 University Ave W, Waterloo, ON N2L 3G1, Canada}
\affiliation{Perimeter Institute for Theoretical Physics, 31 Caroline St. North, Waterloo, ON N2L 2Y5, Canada}
\affiliation{Waterloo Centre for Astrophysics, University of Waterloo, 200 University Ave W, Waterloo, ON N2L 3G1, Canada}
\email{will.percival@uwaterloo.ca}

\author{C.~Poppett}
\affiliation{Lawrence Berkeley National Laboratory, 1 Cyclotron Road, Berkeley, CA 94720, USA}
\affiliation{Space Sciences Laboratory, University of California, Berkeley, 7 Gauss Way, Berkeley, CA  94720, USA}
\affiliation{University of California, Berkeley, 110 Sproul Hall \#5800 Berkeley, CA 94720, USA}
\email{clpoppett@lbl.gov}

\author[0000-0001-7145-8674]{F.~Prada}
\affiliation{Instituto de Astrof\'{i}sica de Andaluc\'{i}a (CSIC), Glorieta de la Astronom\'{i}a, s/n, E-18008 Granada, Spain}
\email{fprada@iaa.es}

\author[0000-0001-6979-0125]{I.~P\'erez-R\`afols}
\affiliation{Departament de F\'isica, EEBE, Universitat Polit\`ecnica de Catalunya, c/Eduard Maristany 10, 08930 Barcelona, Spain}
\email{ignasi.perez.rafols@upc.edu}

\author[0000-0001-7950-7864]{F.~Qin}
\affiliation{Aix Marseille Univ, CNRS/IN2P3, CPPM, Marseille, France}
\email{qin@cppm.in2p3.fr}

\author[0009-0003-4767-9794]{C.~Ross}
\affiliation{School of Mathematics and Physics, University of Queensland, Brisbane, QLD 4072, Australia}
\email{c.ross1@uq.net.au}

\author{G.~Rossi}
\affiliation{Department of Physics and Astronomy, Sejong University, 209 Neungdong-ro, Gwangjin-gu, Seoul 05006, Republic of Korea}
\email{graziano@sejong.ac.kr}

\author[0000-0002-1809-6325]{K.~Said}
\affiliation{School of Mathematics and Physics, University of Queensland, Brisbane, QLD 4072, Australia}
\email{k.saidahmedsoliman@uq.edu.au}

\author[0000-0002-9646-8198]{E.~Sanchez}
\affiliation{CIEMAT, Avenida Complutense 40, E-28040 Madrid, Spain}
\email{eusebio.sanchez@ciemat.es}

\author{D.~Schlegel}
\affiliation{Lawrence Berkeley National Laboratory, 1 Cyclotron Road, Berkeley, CA 94720, USA}
\email{djschlegel@lbl.gov}

\author{M.~Schubnell}
\affiliation{Department of Physics, University of Michigan, 450 Church Street, Ann Arbor, MI 48109, USA}
\email{schubnel@umich.edu}

\author[0000-0002-6588-3508]{H.~Seo}
\affiliation{Department of Physics \& Astronomy, Ohio University, 139 University Terrace, Athens, OH 45701, USA}
\email{seoh@ohio.edu}

\author[0000-0002-3461-0320]{J.~Silber}
\affiliation{Lawrence Berkeley National Laboratory, 1 Cyclotron Road, Berkeley, CA 94720, USA}
\email{jhsilber@lbl.gov}

\author{D.~Sprayberry}
\affiliation{NSF NOIRLab, 950 N. Cherry Ave., Tucson, AZ 85719, USA}
\email{david.sprayberry@noirlab.edu}

\author[0000-0003-1704-0781]{G.~Tarl\'{e}}
\affiliation{University of Michigan, 500 S. State Street, Ann Arbor, MI 48109, USA}
\email{gtarle@umich.edu}

\author[0000-0002-7638-2880]{R. J.~Turner}
\affiliation{Centre for Astrophysics \& Supercomputing, Swinburne University of Technology, P.O. Box 218, Hawthorn, VIC 3122, Australia}
\email{rjturner@swin.edu.au}

\author{B.~A.~Weaver}
\affiliation{NSF NOIRLab, 950 N. Cherry Ave., Tucson, AZ 85719, USA}
\email{benjamin.weaver@noirlab.edu}

\author[0000-0001-5381-4372]{R.~Zhou}
\affiliation{Lawrence Berkeley National Laboratory, 1 Cyclotron Road, Berkeley, CA 94720, USA}
\email{rongpuzhou@lbl.gov}

\author[0000-0002-6684-3997]{H.~Zou}
\affiliation{National Astronomical Observatories, Chinese Academy of Sciences, A20 Datun Road, Chaoyang District, Beijing, 100101, P.~R.~China}
\email{zouhu@nao.cas.cn}

\begin{abstract}
  We calibrate the Tully-Fisher relation (TFR) using observations of spiral 
  galaxies taken during the first year (DR1) of the DESI galaxy redshift survey.  
  The rotational velocities of \Ntot galaxies are measured at $0.4R_{26}$ by 
  comparing the redshifts at $0.4R_{26}$ with those at the galaxy centers of 
  spatially-resolved galaxies targeted as part of the DESI Peculiar Velocity 
  Survey.  The DESI DR1 TFR slope is calibrated by separating the spiral 
  galaxies into redshift bins of width $\Delta z = 0.005$ from $0.03 < z < 0.1$ 
  and jointly fitting the TFR across all bins.  We find a slope of \TFslope in 
  the $r$-band for the TFR, with an intrinsic scatter of \TFscatter.  We present 
  a catalog of the distances and peculiar velocities to these \Ntot galaxies 
  using our calibrated TFR.  For cosmological analyses, we also present a 
  clustering catalog and associated random catalogs using a subset of 6807 of 
  the DESI DR1 TF galaxies.
\end{abstract}

\section{Introduction}\label{sec:intro}

Measuring distances to objects in the universe is a persistent challenge, but 
one that is necessary to tackle to progress the study of most areas of 
astrophysics and cosmology.  Many methods for measuring distances to different 
types of objects have been developed, the most universal of which is Hubble's 
Law \citep{Hubble1929} for extragalactic objects.  While Hubble's Law was almost 
immediately recognized as the ultimate distance indicator due to the ease with 
which one can measure an object's redshift, its distances are plagued at low 
redshifts due to galaxies' peculiar motions contributing to the observed redshift.

The interaction of the galaxy with its local gravitational potential due to the 
matter over- or under-densities produces the peculiar velocity (PV) of the 
galaxy \citep{Davis2014}.  The cosmological redshift of a galaxy, 
$z_{\rm cosmo}$, and that due to its peculiar motion, $z_{\rm pec}$, relate to 
the observed redshift, $z_{\rm obs}$, as \citep{Harrison1974}:
\begin{equation}
    1 + z_{\rm obs} = (1 + z_{\rm cosmo})(1 + z_{\rm pec}).
\end{equation}
Disentangling a galaxy's PV from its cosmological redshift requires a 
redshift-independent distance metric, often in the form of an 
empirically-calibrated distance indicator that relates a distance-independent 
quantity of the galaxy with a distance-dependent property.  Popular 
redshift-independent distance indicators include Type Ia supernovae 
\citep{Phillips1993}, Surface Brightness Fluctuations \citep{Tonry1988}, the 
Fundamental Plane relation \citep[FP;][]{Djorgovski1987,Dressler1987}, and the 
Tully-Fisher relation \citep[TFR;][]{Tully1977}.

The TFR relates the rotational velocity and luminosity of a galaxy.  This 
relation thus applies to spiral galaxies: rotationally-supported systems where 
rotational velocity and luminosity are intrinsically related.  While the 
physical origins of the TFR are not fully understood, a reasonable model for 
their behavior results from Newtonian dynamics.  Assuming that the mass 
distribution of a galaxy is spherically symmetric, the centripetal acceleration 
in an object in circular orbit at radius $r$ is equal to its gravitational 
acceleration from the interior mass of the sphere:
\begin{equation}
    V^2_{\rm rot} \propto \frac{M(r)}{r}.
\end{equation}
In the case of a constant mean surface brightness and mass-to-light ratio for 
spiral galaxies, this relation becomes 
\begin{equation}\label{eqn:TFR_L}
    L \propto V_{\rm rot}^4.
\end{equation}
Many studies have found power-law exponents deviating from 4 both empirically 
\citep[e.g.,][]{Sandage1976, Burstein1982, Bottinelli1983, Pierce1988}, and 
observationally \citep[e.g.,][]{Ponomareva2017, Kourkchi2020a, Douglass2025}.  
These deviations are often attributed to flaws in the assumption of a uniform 
mass-to-light ratio \citep{Mould2020}.

In Tully-Fisher analyses, it is common to determine the rotational velocity of a 
galaxy based on the width of spectral lines such as the 21 cm \ion{H}{1} line 
profile or the H$\alpha$ emission line.  The \ion{H}{1} profile allows for 
perhaps the most direct measurement of the maximum rotational velocity of a 
galaxy when there is sufficient neutral hydrogen present for its detection.  The 
H$\alpha$ emission line is the most prominent in galaxy spectra at visible 
wavelengths, so it is the primary choice as a tracer of the rotational velocity 
of a spiral galaxy in the visible.  It is more difficult to extract the maximum 
rotational velocity of a galaxy using H$\alpha$ emission lines, as measurements 
must be made at sufficiently large radii from the galaxy center to measure a 
flat rotation curve, a region where the H$\alpha$ surface brightness decreases 
rapidly and may be weak \citep{Sofue2001}.  This may result in a systematic 
underestimation of the maximum rotational velocity.

Once determined, the rotational velocity of a galaxy allows the luminosity, or 
absolute magnitude, to be estimated.  The distance modulus is then determined by 
comparing this with the apparent magnitude of the galaxy.  Previous TF distance 
catalogs produced include \cite{Masters2006}, which used a mixture of \ion{H}{1} 
line widths and optical rotation curves to measure the distance to $\sim$5000 
galaxies, and CosmicFlows \citep{Tully2012, Kourkchi2020b}, which relied on 
\ion{H}{1} line widths to measure the distance to $\sim$10,000 galaxies.

In this work, we expand on the TFR analysis performed on the DESI Early Data 
Release described in \citet{Douglass2025}.  We present a catalog of \Ntot PV 
measurements obtained with the TFR using the publicly available DESI Data 
Release 1 (DR1), representing $\sim$20\% of the data expected to be included in 
the final DESI data release.  Alongside this, we present a clustering catalog, 
and associated random catalogs, using a subset of 6807 of these galaxies that 
are used in combination with the DESI DR1 Fundamental Plane sample 
\citep{DESI_DR1_fp} and DESI DR1 mock PV catalogs \citep{DESI_DR1_pv_mocks} in 
the accompanying cosmological analyses of \cite{DESI_DR1_fs8_cf, 
DESI_DR1_fs8_ps, DESI_DR1_fs8_ml}, and \cite{DESI_DR1_zp}.  The TF catalog 
presented here is comparable in size to the TF catalog using \ion{H}{1} line 
widths produced by Cosmicflows-4 \citep[CF4;][]{Kourkchi2020b}, currently the 
largest catalog of distances determined by the TFR.  It covers a much smaller 
area of the sky (3580~sq.~deg. relative to CF4's almost full-sky coverage), but 
it more than doubles the redshift range ($z \lesssim 0.1$ relative to CF4's 
$z < 0.05$) due to the use of the H$\alpha$ emission line instead of \ion{H}{1}.

The layout of this paper is as follows.  We first describe the DESI instrument, 
the DESI Peculiar Velocity Survey, and the DESI Data Release 1 (DR1) in 
Section~\ref{sec:DESI}.  A description of our Tully-Fisher measurements, our 
quality selection criteria, and our estimated systematic uncertainties follow in 
Section~\ref{sec:measure_rot_vel}.  Section~\ref{sec:calibration} discusses the 
calibration of the slope of the Tully-Fisher relation using data from DESI DR1.  
The measured Tully-Fisher relation for \Ntot galaxies in the DESI DR1 is 
presented in Section~\ref{sec:measuring_pv}.  We compare our results to 
previously published TFRs in Section~\ref{sec:discussion}, and we conclude in 
Section~\ref{sec:conclusion}.

When needed, we assume a flat $\Lambda$CDM cosmology with $\Omega_M = 0.3151$ 
and $H_0 = 100h$~km/s/Mpc.

\section{Dark Energy Spectroscopic Instrument}\label{sec:DESI}

A multi-object fiber spectrograph installed on the Mayall 4-meter telescope at 
Kitt Peak National Observatory, the Dark Energy Spectroscopic Instrument 
\citep[DESI;][]{DESI_instrument} is designed to efficiently conduct a 
large-scale redshift survey, collecting $\sim$63 million redshifts over at least 
17,000~sq.~deg. of the sky in 8 years \citep{DESI2016a}.  DESI contains 5000 
robotic fibers covering $\sim 8~{\rm deg}^2$, enabling it to measure up to 5,000 
spectra with a wavelength range of 3600--9800~{\AA} in a single exposure 
\citep{DESI2016b, Silber2023, Miller2024, Poppett2024}.  The instrument was 
designed to allow each fiber to move anywhere within a patrol radius with 
minimal overlap of its neighboring fibers.  DESI's measured spectra are 
processed with 
Redrock\footnote{\href{https://github.com/desihub/redrock}{https://github.com/desihub/redrock}}, 
an offline spectroscopic pipeline using template-matching to extract redshifts 
\citep{Redrock}.

We use the DESI Data Release 1 \citep[DR1;][]{DESI_DR1} in this analysis, which 
contains the spectroscopic observations of 18.7~million objects made during the 
first year of DESI (Y1).  DESI observes objects within one of three programs 
according to the current sky conditions and brightness of the sky (e.g., 
twilight, moon phase, dark sky); the programs are named ``bright,'' ``dark,'' 
and ``backup'' and are essentially treated independently.  Because DESI is 
following a depth-first strategy, DR1 contains 41.3\% of the bright survey 
(9739~sq.~deg.) and and 29.0\% of the dark survey (9528~sq.~deg.).  For more 
details, see \cite{DESI_DR1}.

DESI was designed to make the most precise measurement of the expansion history 
of the Universe to determine the nature of dark energy using Baryonic Acoustic 
Oscillations \citep[BAO;][]{Levi2013, DESI_DR1_BAO, DESI_DR2_BA0}.  To do so, it 
makes use of several target classes/tracers in the main survey, selected from 
the DESI Legacy Imaging Surveys \citep{DESI_Imaging} with the DESI target 
selection pipeline \citep{DESI_target_pipeline}: Bright Galaxies 
\citep[BGS;][]{DESI_target_BGS}, Luminous Red Galaxies 
\citep[LRG;][]{DESI_target_LRG}, Emission Line Galaxies 
\citep[ELG;][]{DESI_target_ELG}, and Quasi-Stellar Objects 
\citep[QSO;][]{DESI_target_QSO}, as well as Milky Way stars 
\citep[MWS;][]{DESI_target_MWS}.  These primary targets often do not make use of 
all 5000 fibers in a single tile \citep[constructed using the optimized survey 
operations pipeline of][]{Schlafly2023}, so DESI fills the spare fibers with a 
number of secondary targeting programs designed by members of the DESI 
collaboration \citep{DESI_target_pipeline}, increasing the scientific output.  
One of these programs is the DESI Peculiar Velocity Survey \citep{Saulder2023}.

\subsection{The DESI Peculiar Velocity Survey}

The DESI Peculiar Velocity (PV) Survey \citep{Saulder2023} is a secondary 
targeting program of DESI that was designed to measure the peculiar velocities 
of galaxies in the local universe ($z \lesssim 0.15$).  The addition of these 
peculiar motions will increase the precision of DESI's measurement of 
$f \sigma_8$, the product of the redshift-dependent growth rate of stucture 
($f$) and the amplitude of the linear power spectrum on the scale of 8~Mpc/$h$ 
($\sigma_8$).  As described in Sec.~\ref{sec:intro}, peculiar velocities can be 
measured when we know the galaxy's observed redshift and its distance through 
some other, redshift-independent, method.  The DESI PV Survey targets galaxies 
whose distances can be measured either via the Tully-Fisher Relation (late-type 
galaxies) or the Fundamental Plane (early-type galaxies).  We focus on the DESI 
DR1 Tully-Fisher sample here; see \cite{DESI_DR1_fp} for a complementary 
discussion of the DESI DR1 Fundamental Plane sample and its calibration, and see 
\cite{Saulder2023} for the DESI PV Survey's target design.

Similar to \cite{Douglass2025}, the late-type galaxies that comprise the TF 
sample are selected from the Siena Galaxy Atlas 2020 \citep[SGA~2020;][]{SGA}.  
Constructed with the DESI Legacy Imaging Surveys DR9 \citep{DESI_Imaging} north 
of $-30^\circ$ declination, the SGA~2020 is a size-limited catalog containing 
galaxies with diameters of the 25th magnitude arcsec$^{-2}$ isophote, $D(25)$, 
greater than 20\arcsec.  Galaxies in the SGA~2020 are included in the TF sample 
if they have a Sersic index $n < 2$ \citep[indicative of a spiral 
galaxy;][]{Blanton2009} and an inclination angle $> 20^\circ$ (an axis ratio, 
$b/a < 0.940$) as measured in the $r$-band.  For each galaxy, the DESI PV Survey 
positions a fiber on the galaxy center and at a distance of $0.4R_{26}$ along 
the semimajor axis of the galaxy on either side of the center; $R_{26}$ is the 
radius of the 26 mag arcsec$^{-2}$ $r$-band isophote reported in the SGA~2020.

\subsection{Photometric corrections}

Similar to \cite{Douglass2025}, we adjust the apparent magnitude of the galaxies 
to account for various systematics.  If $m_r$ is the magnitude within the 26-mag 
arcsec$^{-2}$ isophote in the $r$-band as reported by the SGA-2020, then the 
corrected magnitude, $m_{r, \text{corr}}$ is
\begin{equation}\label{eqn:mag_correct}
  m_{r, \text{corr}} = m_r - A_{\rm MW~dust} - A_{\rm internal~dust} + A_k + A_{\rm sys}.
\end{equation}
Here, $A_{\rm MW~dust}$ is the Milky Way dust attenuation, 
$A_{\rm internal~dust}$ corrects for internal dust extinction, $A_k$ is the 
$k$-correction, and $A_{\rm sys}$ adjusts for the offset between the two 
photometric surveys that comprise the DESI Legacy Surveys.

The extinction correction due to dust in the Milky Way is defined as
\begin{equation}
  A_{\rm MW~dust} = R_r E(B - V),
\end{equation}
where $R_r = 2.165$ and $E(B-V)$ is taken from the updated dust maps by 
\cite{Zhou2024}.  This correction is identical to that used in the DESI EDR TF 
calibration by \cite{Douglass2025}.

\begin{figure}
  \centering
  \includegraphics[width=0.48\textwidth]{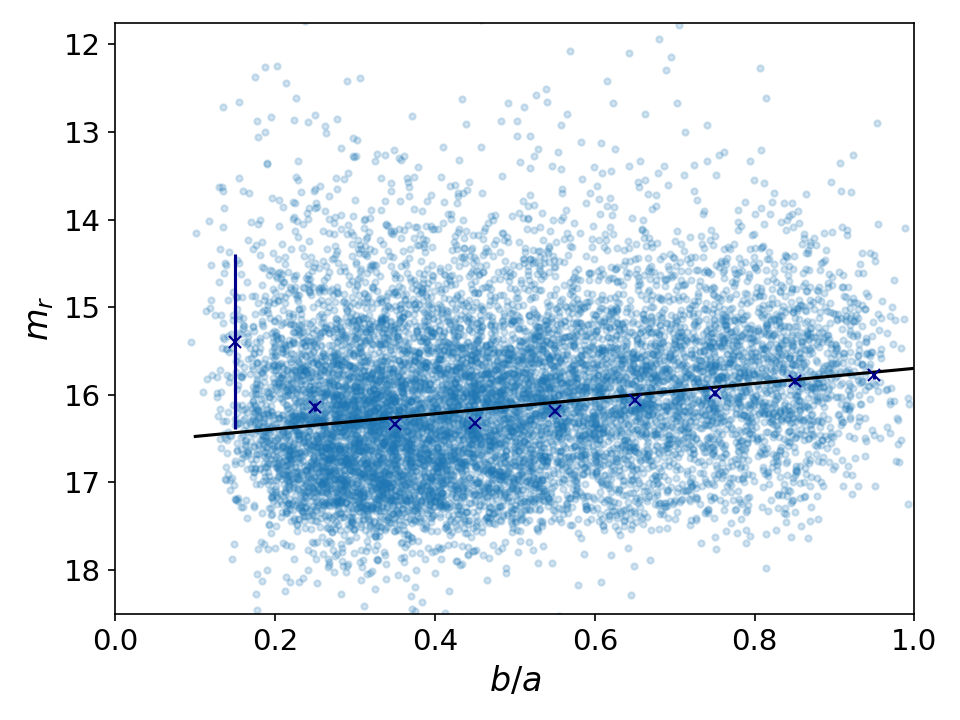}
  \caption{Observed correlation between $r$-band apparent magnitude and axis 
  ratio.  A linear fit to the median magnitudes in each bin (dark blue crosses) 
  is shown in the black line.}
  \label{fig:internal_dust_corr}
\end{figure}

We correct for the extinction due to internal dust within the target galaxy by 
removing the observed correlation between the apparent magnitude and axis ratio 
(shown in Fig.~\ref{fig:internal_dust_corr}).  The correction is defined as 
\begin{equation}
  A_{\rm internal~dust} = d\left( \frac{b}{a} - 1 \right)
\end{equation}
where $d = -0.90^{+0.15}_{-0.18}$ is the slope of the line fit to the median 
magnitudes binned by the axis ratio.  This is an updated correction to that used 
in the DESI EDR TF calibration \cite{Douglass2025}, using the full DR1 TF 
sample.

Identical to \cite{Douglass2025}, we adjust for the zero-point offset reported 
in \cite{Said2024} between the Beijing-Arizona Sky Survey \citep[BASS;][]{BASS} 
and the Mayall $z$-band Legacy Survey \citep[MzLS;][at Dec 
$\geq +32.375^\circ$]{DESI_Imaging} and the Dark Energy Camera Legacy Survey 
\citep[DECaLS;][for Dec $\geq +32.375^\circ$]{DECam}, the two photometric 
surveys that are used in the SGA~2020: 
\begin{equation}
  m_{r, \text{BASS}} - m_{r, \text{DECaLS}} = 0.0234
\end{equation}
with a root mean square (RMS) deviation of 0.02~mag and an error on the mean of 
0.00049.  As in \cite{Douglass2025}, we adjust the BASS magnitudes to shift them 
into the DECaLS frame.

Finally, we $K$-correct the SGA~2020 magnitudes to a redshift of $0.1$ using the 
\texttt{kcorrect} Python package \citep{Blanton2007}.  Note that this is a 
different redshift than was used in the DESI EDR TF calibration 
\citep{Douglass2025}, but matches that used by \cite{DESI_DR1_fp} in the DR1 FP 
calibration.

\section{Measuring the rotational velocity}\label{sec:measure_rot_vel}

\begin{figure}
  \centering
  \includegraphics[width=0.48\textwidth]{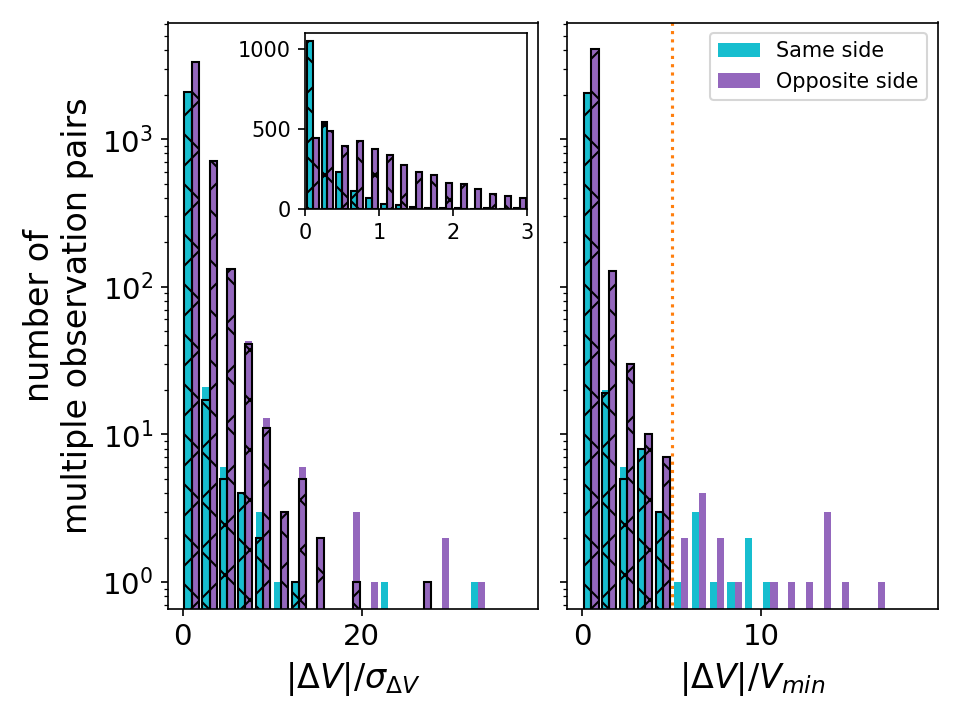}
  \caption{\emph{Left:} Pull distribution of the difference in $V(0.4R_{26})$ 
  for pairs of observations on the same galaxy (on the same or opposite side of 
  the galaxy center).  \emph{Right:} Distribution in $\Delta V / V_{\rm min}$ 
  for galaxies with multiple measures of $V(0.4R_{26})$.  For a given galaxy, 
  all its pairs must fall to the left of the dotted orange line for the galaxy 
  to be in the final sample; those that do are shown in the black hashed 
  histograms.}
  \label{fig:deltaV}
\end{figure}

We compare the redshifts observed at the galaxy's center and at $0.4R_{26}$ to 
estimate the rotational velocity of the galaxy, one of the two quantities in the 
TFR.  For each galaxy, we measure the component of the rotational velocity along 
our line of sight, $V'(0.4R_{26})$, by removing the galaxy's systemic motion 
(the redshift measured at the center of the galaxy) from the measured redshift 
at $0.4R_{26}$:
\begin{equation}\label{eqn:V_from_z}
  \frac{V'(0.4R_{26})}{c} = \frac{1 + z(0.4R_{26})}{1 + z_{\rm center}} - 1
\end{equation}
The redshifts are measured from the spectra using Redrock, DESI's redshift 
pipeline.  Similar to the DESI EDR TF analysis \citep{Douglass2025}, we add a 
7~km/s uncertainty in quadrature to each redshift uncertainty reported by 
Redrock to account for its precision uncertainty.  To recover the tangential 
rotational velocity at $0.4R_{26}$ from the component that we observe along the 
line of sight, we correct each measure of $V'(0.4R_{26})$ with the galaxy's 
inclination angle:
\begin{equation}\label{eqn:Vcorrected}
  V(0.4R_{26}) = \frac{V'(0.4R_{26})}{\sin i}.
\end{equation}
We convert the galaxy's photometric axis ratio from the SGA~2020, $b/a$, to the 
inclination angle as
\begin{equation}\label{eqn:BAtoi}
  \cos^2 i = \frac{(b/a)^2 - q_0^2}{1 - q_0^2},
\end{equation}
with $q_0 = 0.2$ \citep{Tully2000}.

We use the same quality criteria as with the DESI EDR TF calibration 
\citep{Douglass2025}:
\begin{itemize}
  \item Center redshifts must have \texttt{DELTACHI2} $> 25$ and \texttt{ZWARN} 
        $= 0$;
  \item Measured rotational velocities $V(0.4R_{26})$ must be between 
        10--1000~km/s;
  \item When there are multiple measures of $V(0.4R_{26})$ for a given galaxy, 
        only rotational velocities with \texttt{DELTACHI2} $> 25$ that also 
        satisfy $\Delta V / V_{\rm min} < 5$ are used, where $\Delta V$ is the 
        difference between any two of the $V(0.4R_{26})$ and $V_{\rm min}$ is 
        the minimum $V(0.4R_{26})$ in that galaxy.
\end{itemize}
Uncertainties in $V(0.4R_{26})$ are estimated by propagating the redshifts 
uncertainties through Eqns.~\ref{eqn:V_from_z}--\ref{eqn:Vcorrected}.  
Fig.~\ref{fig:deltaV} shows the pull distribution for $\Delta V$ (on the left) 
and $\Delta V / V_{\rm min}$ (on the right).  As the inset on the left shows, 
our estimates on the velocity uncertainties are appropriate.

If a galaxy has multiple measures of $V(0.4R_{26})$ that pass all quality cuts, 
we use the weighted average of all the $V(0.4R_{26})$ as that galaxy's 
rotational velocity.

\subsection{Rotational velocity systematics}

To confirm the accuracy with which we are recovering the rotational velocity 
with a single measurement along a galaxy's major axis, we compare the velocities 
at $0.4R_{26}$ as measured in the DESI PV Survey to the rotational velocities at 
$0.4R_{26}$ based on models of H$\alpha$ velocity fields from the SDSS Mapping 
Nearby Galaxies at APO Survey \citep[MaNGA]{MaNGA}.  We investigate the 
differences in velocities for 207 galaxies that are in the DR1 TF sample and are 
observed in SDSS MaNGA DR17 and modeled by \cite{Ravi2023}, in order to evaluate 
the systematics associated with measuring velocities at $0.4R_{26}$.

\subsubsection{SDSS MaNGA}

SDSS MaNGA DR17 \citep{SDSS_DR17} was the final data relase of MaNGA, an 
integral field spectroscopy survey of $\sim$10,000 galaxies.  The MaNGA integral 
field units (IFUs) contain between 19 and 127 fibers covering {12.5\arcsec} to 
{32.5\arcsec} in diameter.  The light received by the IFU was fed to two 
spectrographs with wavelength ranges of 3600--10,300~{\AA} with a resolution of 
$\lambda/\Delta \lambda \sim 2000$ \citep{Smee2013}.  The SDSS MaNGA sample 
contains galaxies observed out to $1.5R_e$ or $2.5R_e$, where $R_e$ is the 
half-light radius of a galaxy.  Using two-dimensional H$\alpha$ velocity fields 
from SDSS MaNGA, \cite{Ravi2023} model the rotational velocity of $\sim$5500 
disk galaxies. 

\cite{Ravi2023} fit the H$\alpha$ velocity maps from SDSS MaNGA to the following 
rotation curve parameterization \citep{BarreraBallesteros2018}:
\begin{equation}\label{eqn:MaNGA_rot_curve}
  V(r) = \frac{V_\text{max}r}{(R_\text{turn}^\alpha + r^\alpha)^{1/\alpha}}.
\end{equation}
$V(r)$ is the rotational velocity  at distance $r$ from the center of each 
galaxy.  The parameterization describes a rotation curve where the velocity 
increases until distance $R_{\rm turn}$ from the center of the galaxy, turns 
with sharpness $\alpha$, and plateaus at velocity $V_{\rm max}$.  $V_{\rm max}$, 
$R_{\rm turn}$, $\alpha$, as well as the inclination angle $i$, the kinematic 
position angle $\phi$, and the kinematic center of the galaxy are free 
parameters in the fit. See \cite{Ravi2023} for details of the modeling 
procedure.

\subsubsection{Kinematic v. photometric position angle}

\begin{figure}
  \centering
  \includegraphics[width=0.48\textwidth]{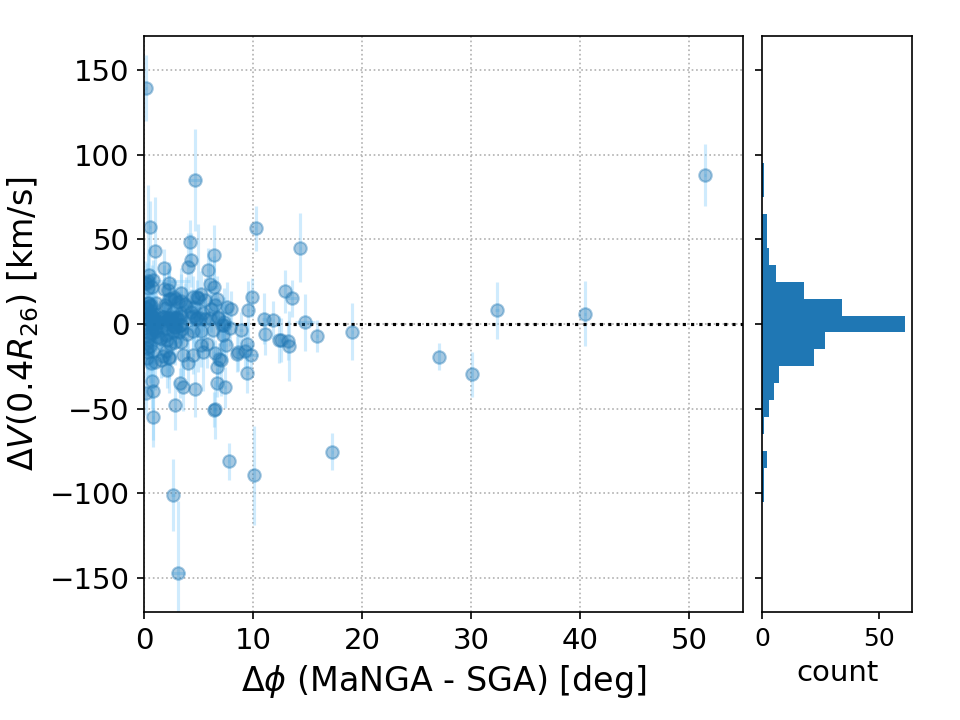}
  \caption{Difference in the velocity at $0.4R_{26}$ as measured in the DESI PV 
  Survey and calculated from the SDSS~MaNGA velocity field fits from 
  \cite{Ravi2023} as a function of the difference in the photometric position 
  angle from the SGA and the kinematic position angle from the velocity field 
  fits from \cite{Ravi2023}.}
  \label{fig:dV_dPhi}
\end{figure}

We investigate if the velocity measurements at $0.4R_{26}$ are affected by 
incorrect position angles by examining the difference between the velocity at 
$0.4R_{26}$ measured in the DESI PV survey and modeled by \cite{Ravi2023} as a 
function of the difference in the photometric position angle used for targeting 
in the DESI PV survey and the kinematic position angle from \cite{Ravi2023}.

The DESI PV Survey uses the SGA's photometric position angles $\phi$, defined as 
the position of a galaxy's major axis east of north, to place a DESI fiber at 
galactocentric radius $0.4R_{26}$ along a galaxy's major axis.  The rotational 
velocity calculated with Eqn.~\ref{eqn:Vcorrected} needs to be corrected for the 
difference between the kinematic and photometric position angles.

In Fig.~\ref{fig:dV_dPhi}, we show the difference between the rotational 
velocity measurements at $0.4R_{26}$ from the DESI PV Survey and the rotational 
velocities at $0.4R_{26}$ obtained from the SDSS~MaNGA velocity field modeling, 
$\Delta V(0.4R_{26})$, as a function of the difference between the photometric 
position angle from the SGA and the kinematic position angle from the fit by 
\cite{Ravi2023}, $\Delta \phi$.  We see no correlation between the difference in 
position angles and the difference in velocities.  There are three galaxies with 
$\Delta V(0.4R_{26}) > 100$~km/s.  Of these three, one galaxy has a velocity map 
that appears to be affected by the presence of an AGN.  The other two galaxies 
are small in angular size compared with the size of the DESI fiber, which may 
affect the velocity measurement.

\subsubsection{Velocity dispersion introduced with DESI}

\begin{figure*}
  \includegraphics[width=0.49\textwidth]{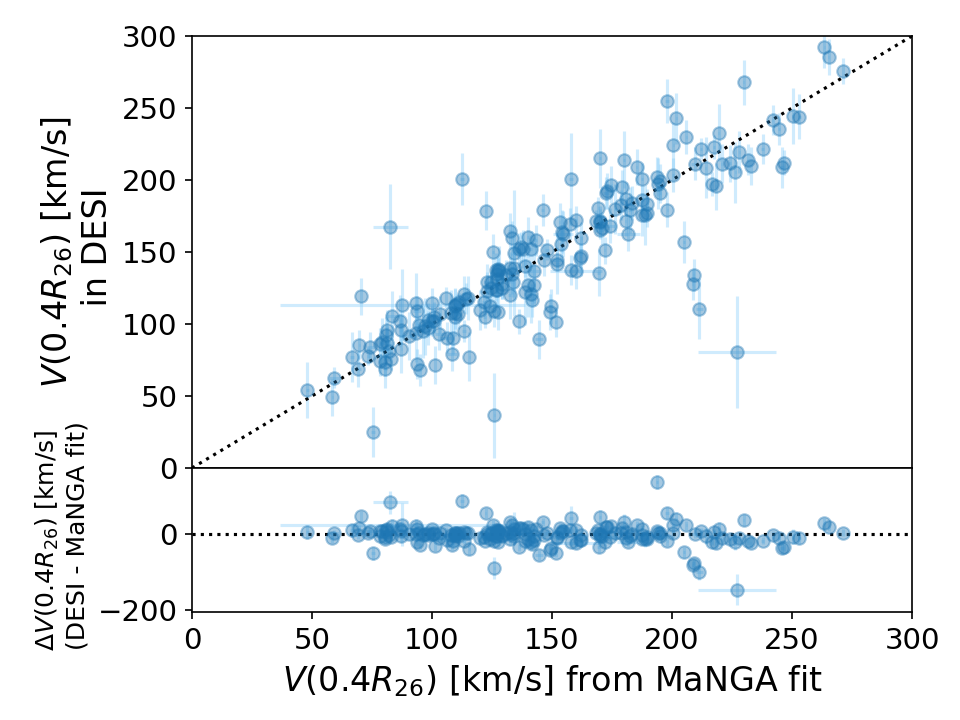}
  \includegraphics[width=0.49\textwidth]{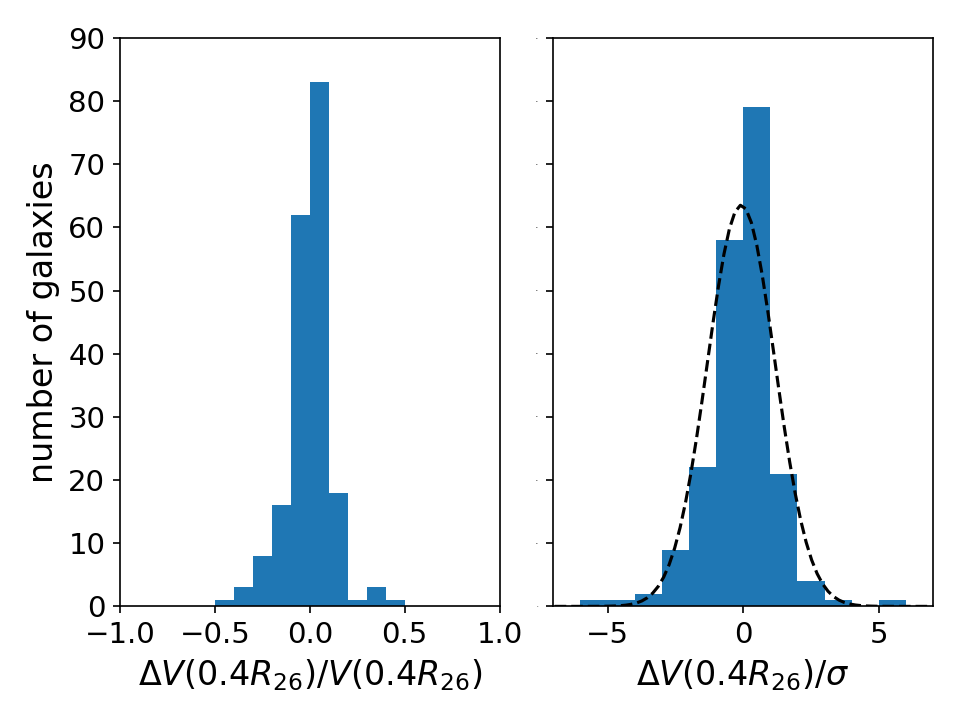}
  \caption{Rotational velocities at $0.4R_{26}$ as measured in the DESI PV 
  Survey and calculated from the the SDSS~MaNGA velocity field fits from 
  \cite{Ravi2023} corrected for the differences in inclination and position 
  angle between the two measurements.  The left figure compares the two 
  velocities with the black dotted line indicating where the velocities are 
  equal.  The center histogram shows the perpendicular distance of each galaxy 
  to $y = x$, $\Delta V(0.4R_{26})$, normalized by the average of the two 
  velocity measurements, the expected value on $y = x$.  The right histogram 
  shows the pull distribution of $\Delta V(0.4R_{26})$.  The black dashed line 
  shown a Gaussian fit to the pull distribution with a mean of $-0.07$ and 
  standard deviation of 1.25.}
  \label{fig:V0p4R26_MaNGA_comparison}
\end{figure*}

We can compare the observed velocity at $0.4R_{26}$ in the DESI PV Survey to the 
predicted velocity at $0.4R_{26}$ from the rotation curve fits of 
\cite{Ravi2023} for galaxies observed in both the DESI PV Survey and SDSS MaNGA 
in order to understand the the dispersion due to measuring the velocity at a 
single radius rather than over the face of each galaxy. 

To remove the differences in velocities due to differences in the position angle 
and inclination angle between the SGA photometric fits and the best-fit values 
of \cite{Ravi2023}, we first correct the predicted velocities at $0.4R_{26}$, 
$V_{\rm MaNGA}$, from the SDSS~MaNGA rotation curve fits.  The corrected 
velocity $V_{\rm MaNGA, expected}$, the velocity at $0.4R_{26}$ from the 
rotation curve fits using the orientation of the galaxy provided by the SGA, is 
calculated as follows. 
\begin{equation}
  V_\text{MaNGA, expected} = A(Q \sin i) B(\Delta \phi) V_\text{MaNGA},
\end{equation}
where
\begin{equation}
  A(Q \sin i) = \frac{\sin i_\text{MaNGA}}{\sin i_\text{SGA}},
\end{equation}
and
\begin{equation}
  B(\Delta \phi) = \cos (|\phi_\text{MaNGA} - \phi_\text{SGA}|).
\end{equation}
Here, $i_{\rm MaNGA}$ and $\phi_{\rm MaNGA}$ are  the best-fit inclination 
position angle from \cite{Ravi2023} respectively.  The photometric inclination 
angle and position angle from the SGA are denotes as $i_{\rm SGA}$ and 
$\phi_{\rm SGA}$. 

In Fig.~\ref{fig:V0p4R26_MaNGA_comparison}, we compare the velocity measured at 
$0.4R_{26}$ in the DESI PV Survey and the expected velocity from the SDSS MaNGA 
rotation curve fits.  The velocities lie along the $y =x$ line with some scatter 
in the relation.  We measure this scatter by calculating the perpendicular 
distance of each point to the $y=x$ line.  The center figure of 
Fig.~\ref{fig:V0p4R26_MaNGA_comparison} shows the distribution of these 
distances normalized by the expected position on the $y=x$ line.  The standard 
deviation of this distribution is 0.15.

\section{Calibrating the Tully-Fisher Relation}\label{sec:calibration}

We calibrate the TFR (Eqn.~\ref{eqn:TFR_L}) in the form
\begin{equation}\label{eqn:TFR}
  M_r = a\log \left( \frac{V}{V_0} \right) + b
\end{equation}
where $M_r$ is the absolute magnitude of the galaxy in the $r$-band within the 
26-mag isophote, $\log V_0 = 2.18$ is the median value of $\log V(0.4R_{26})$ of 
the subset of galaxies used in the calibration, $a$ is the slope of the TFR, and 
$b$ is the zero-point.  As done in \cite{Douglass2025} with the DESI EDR TF 
relation, we use a non-zero value for $\log V_0$ to minimize the correlation 
between the calibrated slope and $y$-intercepts of the TFR.

Similar to \cite{Douglass2025}, we calibrate the TFR using a subset of the full 
DR1 TF sample.  To ensure that we have a reasonable estimate of $V(0.4R_{26})$, 
we require all calibration galaxies to:
\begin{itemize}
  \item Have an inclination angle $i > 45^\circ$,
  \item Have a rotational velocity 
        $70\text{ km/s} < V(0.4R_{26}) < 300\text{ km/s}$ and 
        $\log V(0.4R_{26}) < 0.3(\mu(z_{\rm CMB}) - 34 - 5\log h) + 2$,
  \item Have an apparent magnitude $m_r$ larger than the minimum of 
        $(17.75, \mu(z_{\rm CMB}) - 17 + 5\log h)$,
  \item Have a spiral morphology (described in detail in 
        Sec.~\ref{sec:morphology} below), and 
  \item Pass visual inspection,
\end{itemize}
where $\mu(z_{\rm CMB})$ is the distance modulus based off the observed DESI 
redshift in the CMB frame.  Introducing quality cuts that depend on an assumed 
cosmology does not affect our calibration because, as described in 
Sec.~\ref{sec:define_zbins}, we calibrate in redshift bins, so these criteria 
translate to a uniform cut in each bin that is consistent across all bins.  The 
visual inspection is used to, for example, remove galaxies with either obvious 
tidal features or are overlapping another galaxy, or observations which are 
close to a bright star.

\subsection{Identifying galaxy morphology}\label{sec:morphology}

\begin{figure*}
  \includegraphics[width=\textwidth]{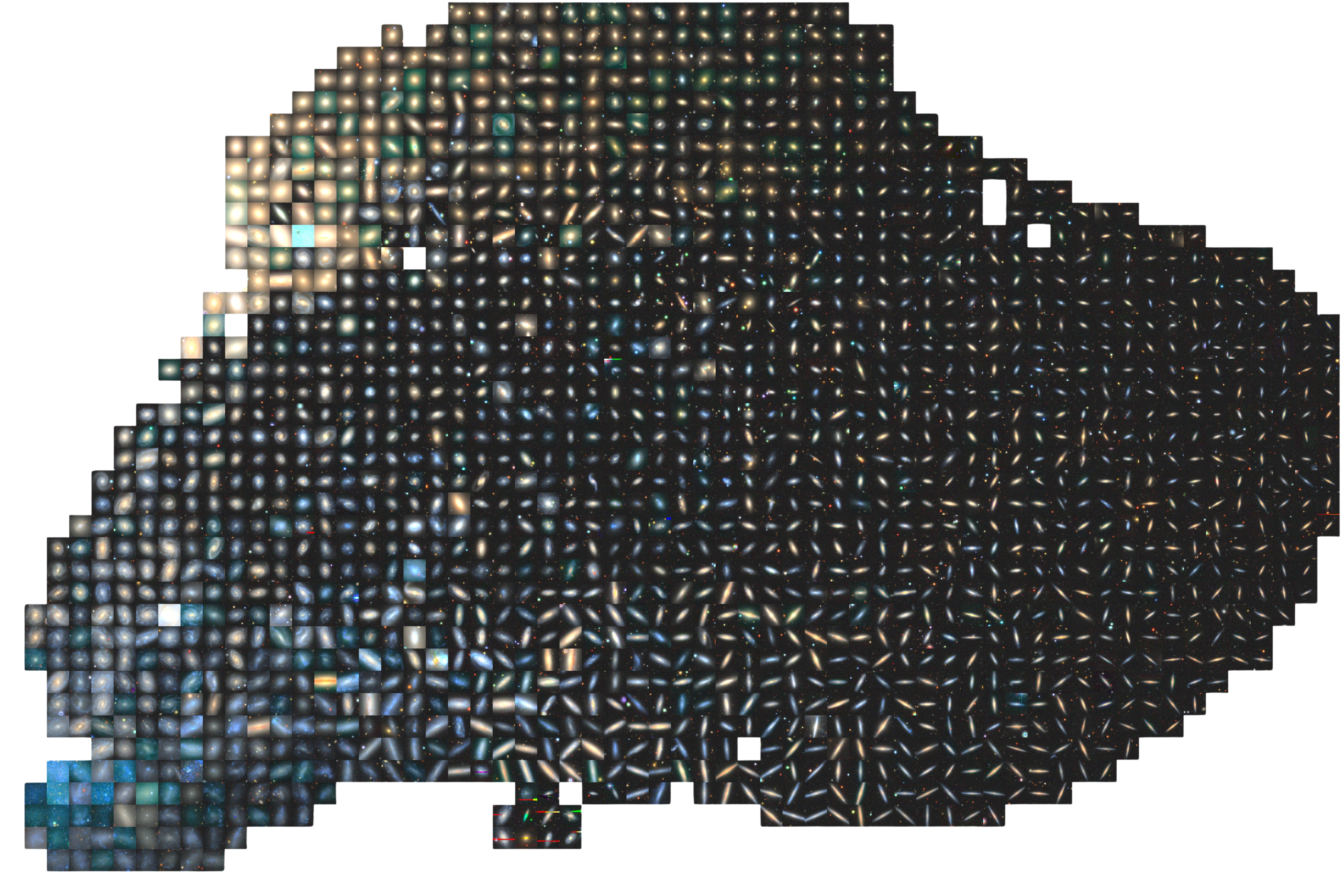}
  \includegraphics[width=0.48\textwidth]{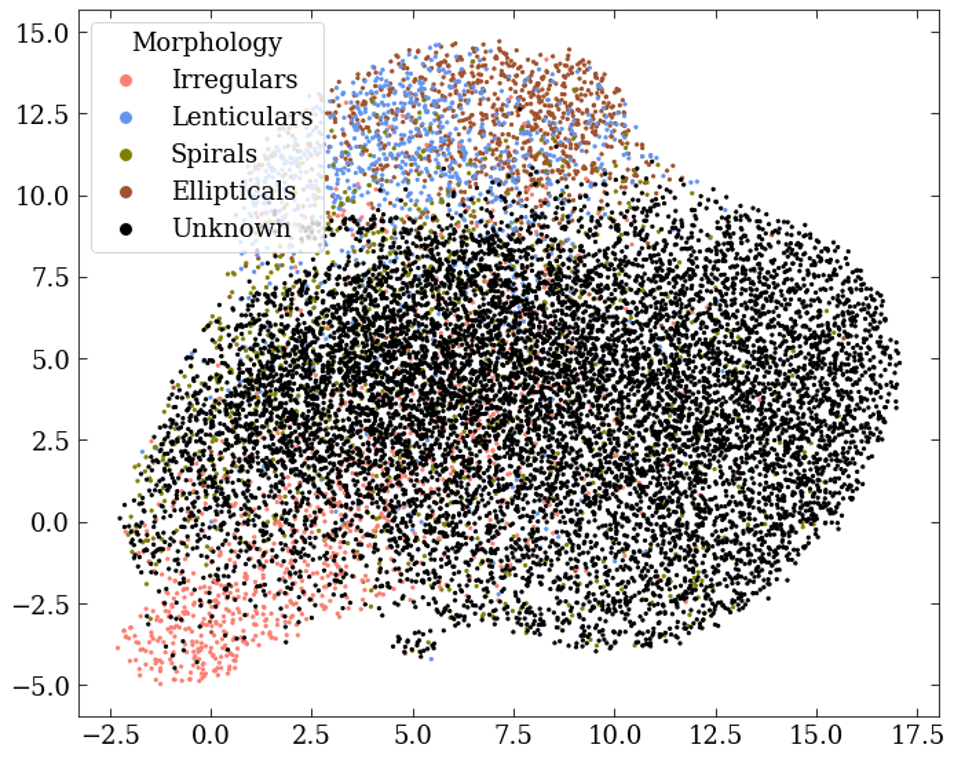}
  \includegraphics[width=0.48\textwidth]{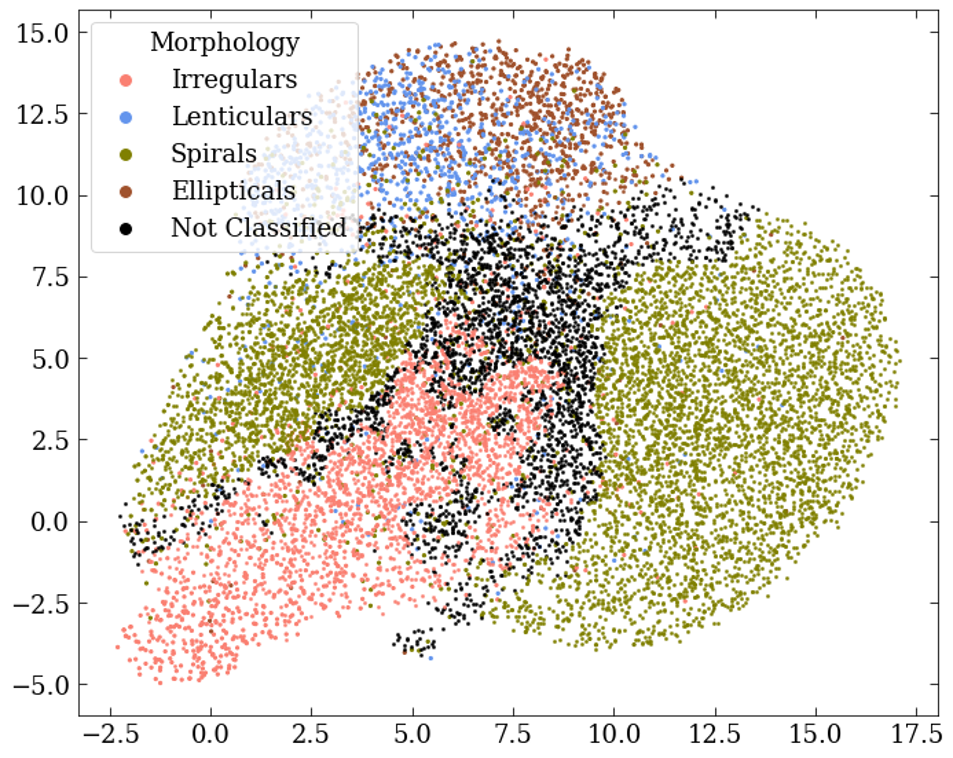}
  \caption{Uniform Manifold Approximation and Projection (UMAP) of the DESI DR1 
  TF sample.  \emph{Top:} We show the cutouts of a random subset of 1,718 
  galaxies from the full sample of 10,262 DESI DR1 TF galaxies + 4000 anchor 
  galaxies, to demonstrate how the classifier sorts the galaxies by similar 
  appearance.  This subset of galaxies are a mix of galaxies with unknown 
  classifications and anchor galaxies of Elliptical, Spiral, Lenticular, or 
  Irregular classifications, as shown in the bottom left plot.  \emph{Bottom 
  left:} Complete UMAP of the full 10,262 DESI DR1 TF sample (shown as black 
  points) along with the 4000 anchor galaxies with known morphological 
  classifications (Irregulars in pink, Lenticulars in blue, spirals in olive, 
  and ellipticals in brown).  A random subset of this projection is shown on top 
  with cutouts of the galaxy images to show how the classifier shorts the images 
  based on their appearance.  We see that the anchor ellipticals are found 
  grouped together towards the top right, anchor lenticulars in the top left, 
  anchor irregulars in the bottom left, and anchor spirals are spread across the 
  middle and right.  The galaxies being classified (black points) sit primarily 
  across the central and right regions of this projection, where the anchor 
  spirals also sit, so the vast majority of the DESI DR1 TF sample are spiral 
  galaxies.  \emph{Bottom right:} Result after the first classification 
  iteration, showing the anchor galaxies and the newly classified galaxies in 
  their corresponding colors.  Those galaxies which have multiple potential 
  classifications (because they sit in the overlap regions between the anchor 
  points) are shown in black and will be evaluated in the second iteration.}
  \label{fig:tfy1_umap}
\end{figure*}

As the TFR is applicable only to galaxies supported against gravitational 
collapse by rotation (usually late-type spiral galaxies), we apply machine 
learning to develop an automated morphological classification system for the 
DESI DR1 TF galaxy sample to identify spiral galaxies.  Using imaging from the 
DESI Legacy Surveys Data Release 9 \citep[DR9;][]{DESI_Imaging}, we adapt the 
pre-trained self-supervised model \texttt{ssl\_legacysurvey} 
\citep{stein2021self}, which embeds galaxies in a high-dimensional latent space 
based on their color and morphology.  The similarity search algorithm sorts 
images based on representations created from multiple augmentations of the 
original image, allowing the model to learn different morphological features.  
These learned representations are then projected to two dimensions using a 
Uniform Manifold Approximation and Projection \citep[UMAP;][]{mcinnes2018umap}, 
as shown in Fig.~\ref{fig:tfy1_umap}, to visualize morphological similarity.

To assign morphological classes to each galaxy, we construct a labeled anchor 
sample of 4,000 galaxies from the SGA 2020, consisting of 1,000 
visually-inspected galaxies for each of the following morphological types: 
Spiral, Elliptical, Lenticular, and Irregular.  Galaxies were randomly selected 
from the SGA sample that had existing classifications in the HyperLeda catalog 
\citep{makarov2014}.  We visually inspected each selected galaxy to confirm its 
classification, building up an anchor sample of 1,000 galaxies of each type.  
During visual inspection, we use qualitative morphological criteria: 
\begin{description}
  \item[Spiral] The presence of a disk with visible spiral arms, and often a 
        central bulge;
  \item[Elliptical] Smooth centrally concentrated light profiles with round or 
        ellipsoidal shapes, and no visible spiral arms;
  \item[Lenticular] Smooth disk that lacks spiral arms and has a prominent 
        central bulge;
  \item[Irregular] Disturbed asymmetric structure without a clear disk or bulge.
\end{description}

We then project the \Ntot unclassified galaxies from the DESI DR1 TF sample into 
the same UMAP space, shown in Fig.~\ref{fig:tfy1_umap}.  Since morphological 
labels are assigned prior to the VI-based quality cuts, some of the UMAP cutouts 
may include bright stars or tidal features.  We note that dwarf galaxies may 
also be present, but are removed from the final sample, as described in 
Sec.~\ref{sec:Alex_cuts}; the anchor galaxies were manually screened to avoid 
these cases.  Next, we apply four separate binary $k$-nearest neighbor 
classifiers \citep[k-NN;][]{cover1967nearest}, each trained to recognize one of 
the four morphological types.  Each k-NN classifier evaluates whether a galaxy 
belonged to its target class by identifying the most similar labeled galaxies in 
the embedding and assigning a label based on the majority class among those 
neighbors.

We combine the four predictions to identify cases of agreement.  Galaxies that 
receive a positive prediction from exactly one classifier were confidently 
assigned that morphological type, while ambiguous cases (those receiving 
multiple or no positive predictions) are left unlabeled.  To improve coverage, 
we implement an iterative process.  After each classification round, the 
remaining unclassified galaxies are re-projected using a new UMAP embedding 
alongside the fixed anchor set, and the k-NN classification is repeated.  We 
performed three iterations in this analysis, with 5\% remaining unclassified, 
typically due to their location near morphological boundaries in the embedding 
space.

Using this method, we achieve a mean classification accuracy of 82\%, estimated 
via 5-fold cross-validation against the anchor point labels, and a completeness 
of 95\% for the DESI DR1 TF sample.  As described in the next section, galaxies 
must be positively identified as spirals with this classification technique to 
be used to calibrate the TFR.  As described in \cite{Saulder2023}, our sample 
selection for galaxies in the TF sample of the DESI PV Survey was designed to 
preferentially select spiral galaxies, so most of our sample is late-type 
spirals.  However, there are some interloping lenticular galaxies, as well as 
some ellipticals and irregular galaxies.  These comprise a relatively small 
fraction of the full DESI Y1 TF sample of \Ntot galaxies; we provide the SSL 
morphology classifications if a science application requires a pure spiral 
galaxy sample.

\subsection{Defining galaxy samples for calibration}\label{sec:define_zbins}

The TFR can be calibrated using either apparent or absolute magnitudes.  To 
calibrate using apparent magnitudes requires knowing or assuming a distance to 
the galaxies \emph{a priori}, severely limiting the sample size.  Using the 
apparent magnitudes typically allows a larger sample of the galaxies to be used 
for calibration, but it requires grouping the galaxies so that all galaxies 
within a given subset are the same distance from us.  This is typically done by 
assigning galaxies to clusters \citep[e.g.,][]{Kourkchi2020a, Douglass2025}, 
which requires a reliable cluster catalog and assumes that the TFR is the same 
in clusters as in the field \citep[see, e.g.,][for studies assessing the 
accuracy of this assumption]{MilvangJensen2003, Ziegler2003, Bamford2005, 
Nakamura2006, Mocz2012}.

To avoid these requirements and use the largest possible set of galaxies for 
calibrating, we separate the galaxies into redshift bins of width 
$\Delta z = 0.005$ from $0.03 \leq z_{\rm CMB} \leq 0.1$.  The lower limit of 
our redshift range is chosen to avoid calibrating with galaxies whose peculiar 
velocities potentially comprise a significant fraction of their observed 
redshifts, and the upper limit is selected based on the maximum redshift used 
for the associated cosmological analysis.  There are a total of 4050 galaxies 
among these 14 redshift bins, with between 116 and 356 galaxies in each bin.  
See Table~\ref{tab:zbin_counts} for the number of galaxies in each redshift bin.

\begin{deluxetable}{ccc}
  \tablewidth{0pt}
  \tablecaption{Number of calibration galaxies in each redshift bin\label{tab:zbin_counts}}
  \tablehead{\colhead{Redshift range} & \colhead{Number} & \colhead{rms scatter} \\ [-0.5em] & \colhead{of galaxies} & \colhead{[AB mag]}}
  \startdata
    0.030--0.035 & 338 & 0.55\\
    0.035--0.040 & 309 & 0.60\\
    0.040--0.045 & 323 & 0.56\\
    0.045--0.050 & 350 & 0.60\\
    0.050--0.055 & 356 & 0.60\\
    0.055--0.060 & 312 & 0.57\\
    0.060--0.065 & 328 & 0.61\\
    0.065--0.070 & 317 & 0.60\\
    0.070--0.075 & 294 & 0.56\\
    0.075--0.080 & 297 & 0.61\\
    0.080--0.085 & 303 & 0.59\\
    0.085--0.090 & 242 & 0.50\\
    0.090--0.095 & 165 & 0.66\\
    0.095--0.100 & 116 & 0.46
  \enddata
\end{deluxetable}

\subsection{Joint calibration of the Tully-Fisher Relation}\label{sec:joint_cal}

\begin{figure*}
  \centering
  \includegraphics[width=\textwidth]{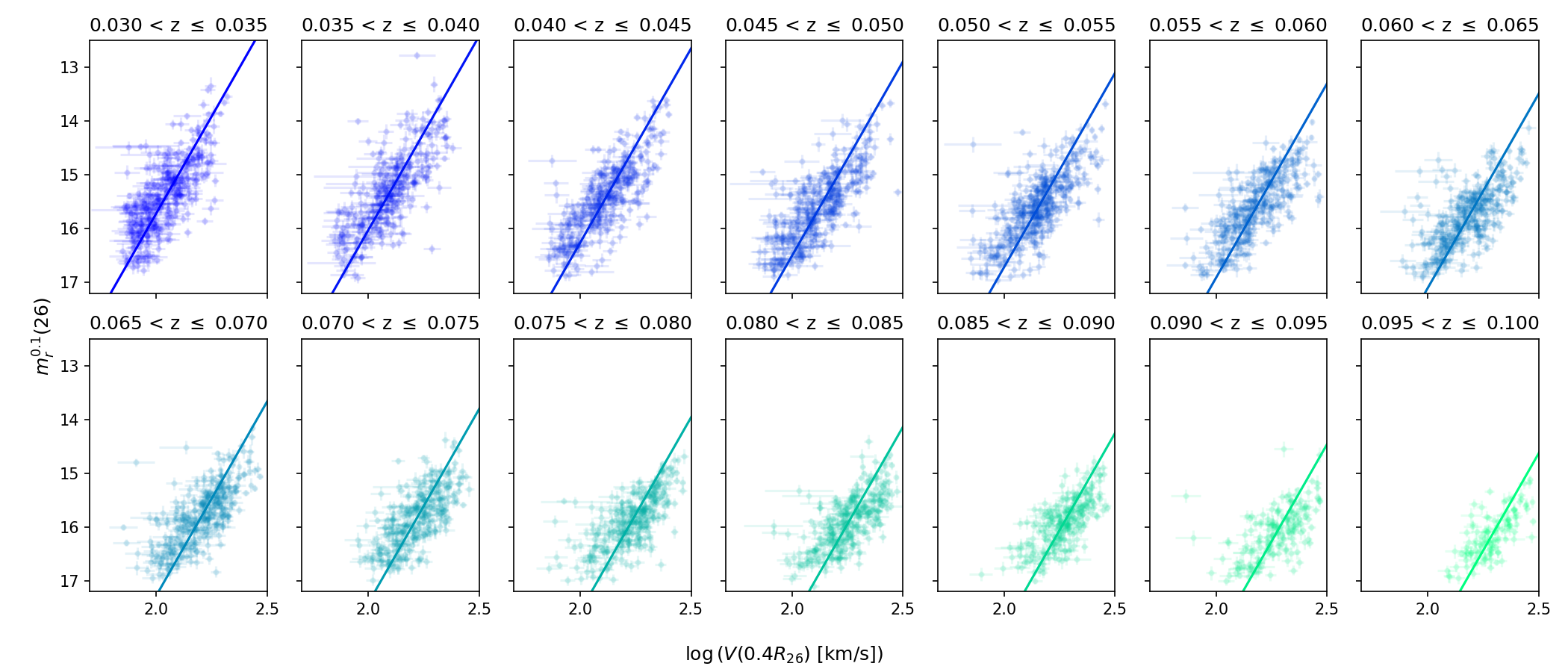}
  \caption{The 4050 DESI DR1 galaxies used for calibrating the TFR, shown in 
  redshift bins $0.03 < z_{\rm CMB} < 0.1$ of width $\Delta z = 0.005$.  The 
  calibrated TFR for each redshift bin is drawn with the colored solid lines.  
  The slope of the TFR is the same across all bins, but each redshift bin has 
  its own $y$-intercept.  The apparent magnitudes used for calibration are from 
  Eqn.~\ref{eqn:mag_correct}, which have been $K$-corrected to $z = 0.1$.}
  \label{fig:TFR_cal}
\end{figure*}

We perform a joint fit of the galaxies in each of the redshift bins, requiring 
all bins to share the same slope, but each to have its own $y$-intercept.  (The 
$y$-intercepts will vary from bin to bin as they are in apparent-magnitude 
space, which is a function of distance.)  We use the subclassed 
HyperFit\footnote{Available at \url{https://github.com/CullanHowlett/HyperFit}} 
package \citep{Robotham2015} that was used in \cite{Douglass2025} to perform the 
fit so that we can account for the residual scatter in the TFR.  We are fitting 
a hybrid of both the TFR and inverse TFR, as HyperFit minimizes the distance 
perpendicular to the best-fit line:
\begin{equation}\label{eqn:tfr_bins}
  m_r = a\log \left( \frac{V}{V_0} \right) + b_i,
\end{equation}
where $b_i$ is the $y$-intercept for the $i^{\rm th}$ redshift bin, by 
maximizing the log-likelihood
\begin{multline}
  \ln \mathcal{L} = \frac{1}{2} \sum_{i=1}^{14} \sum_{j=1}^{N_i} \left[ \ln \left( \frac{a^2 + 1}{s_j^2} \right) \right. \\ \left. - \frac{\left(a\log \left( \frac{V_j}{V_0} \right) + b_i - m_{r,j} \right)^2}{s_j^2} \right],
\end{multline}
where $s_j^2 = \sigma^2 + \sigma_{\log(V_j)}^2 a^2 + \sigma_{m_{r,j}}^2$ and 
$N_i$ is the number of galaxies in the $i^{\rm th}$ redshift bin.  We have 16 
free parameters: the slope $a$, the $y$-intercepts $b_{1\cdots14}$ for each of 
the 14 redshift bins, and the intrinsic scatter of the TFR, $\sigma$.

\begin{figure*}
  \centering
  \includegraphics[width=\textwidth]{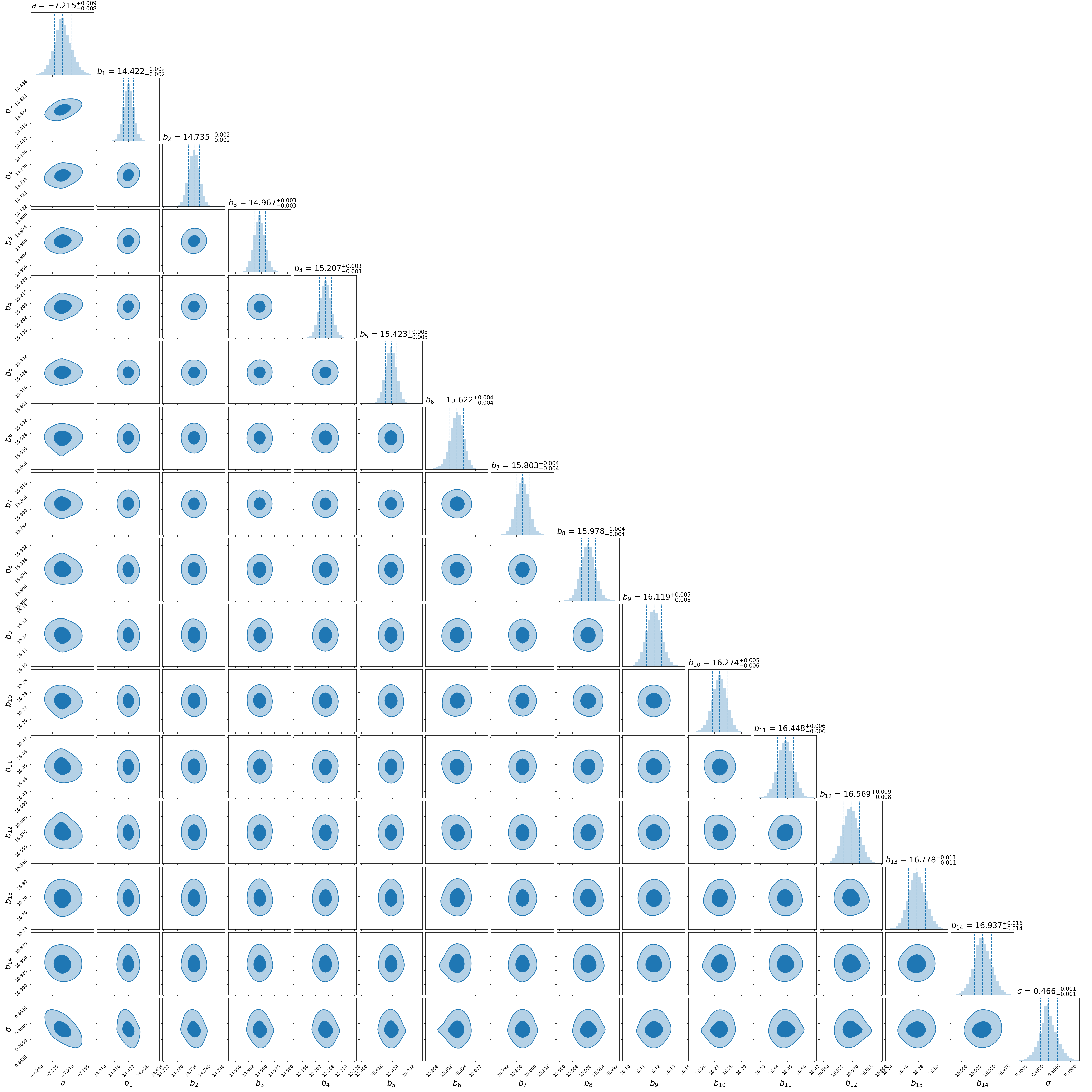}
  \caption{Corner plot of the linear fits to the galaxies in the redshift bins 
  shown in Fig.~\ref{fig:TFR_cal}, where $a$ is the slope, $b_i$ for 
  $i = 1\ldots14$ is the $y$-intercept of the each bin's TFR, and $\sigma$ is 
  the intrinsic scatter along the magnitude axis.}
  \label{fig:TFR_corner}
\end{figure*}

Fig.~\ref{fig:TFR_cal} shows the TFR for our 4050 calibration galaxies in the 14 
redshift bins, with a best-fit slope value of \TFslope.  The intrinsic scatter 
along the magnitude axis for the sample of calibration galaxies is \TFscatter, 
comparable to that found in earlier TF calibrations \citep[e.g., 0.41~mag and 
0.49~mag in][respectively]{Tully2012, Kourkchi2020a}.  The corner plot for the 
TFR calibration is shown in Fig.~\ref{fig:TFR_corner}.

\section{Measuring Peculiar Velocities}\label{sec:measuring_pv}

\begin{figure}
  \centering
  \includegraphics[width=0.45\textwidth]{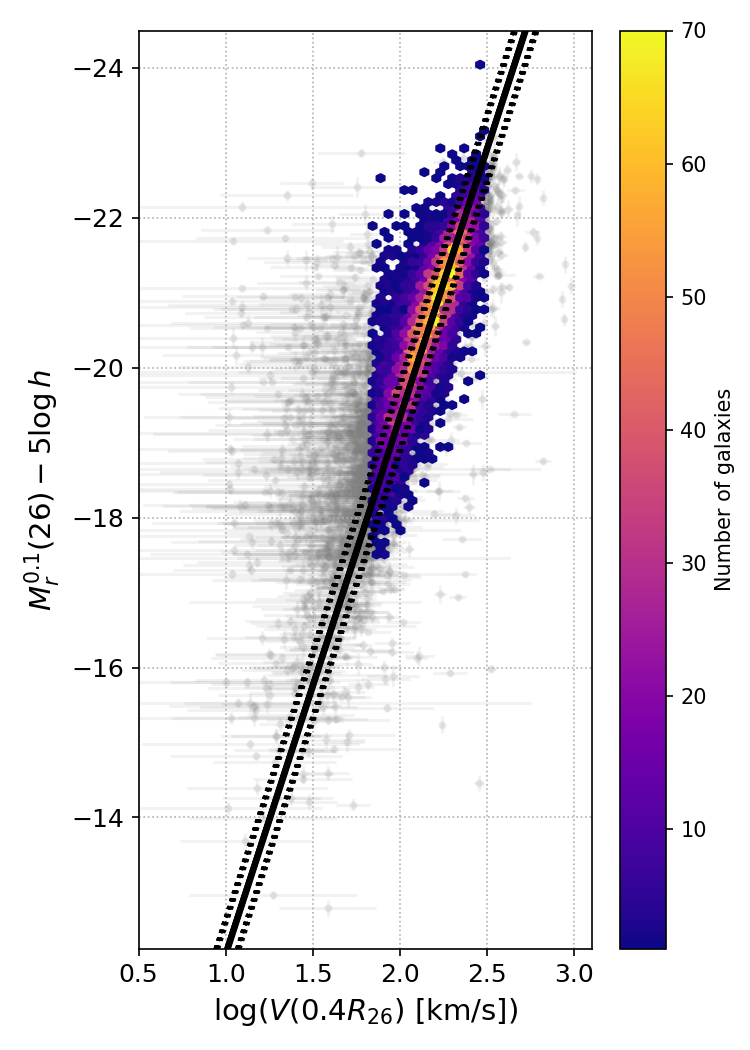}
  \caption{TFR of the DESI DR1 PV Survey galaxies, using the corrected absolute 
  magnitudes described by Eqn.~\ref{eqn:mag_correct} that have been 
  $k$-corrected to $z = 0.1$.  Our calibrated TFRs are shown in the thick solid 
  black line, and the intrinsic dispersion of the calibrated TFR is shown with 
  the black dotted lines.  Galaxies shown in gray are those which are not part 
  of the main cosmological sample, due either to being considered dwarfs or not 
  satisfying the velocity criteria described in Sec.~\ref{sec:Alex_cuts}.  
  Galaxies are scattered vertically in this diagram due to their peculiar 
  motion.}
  \label{fig:TFR}
\end{figure}

\begin{figure}
  \centering
  \includegraphics[width=0.45\textwidth]{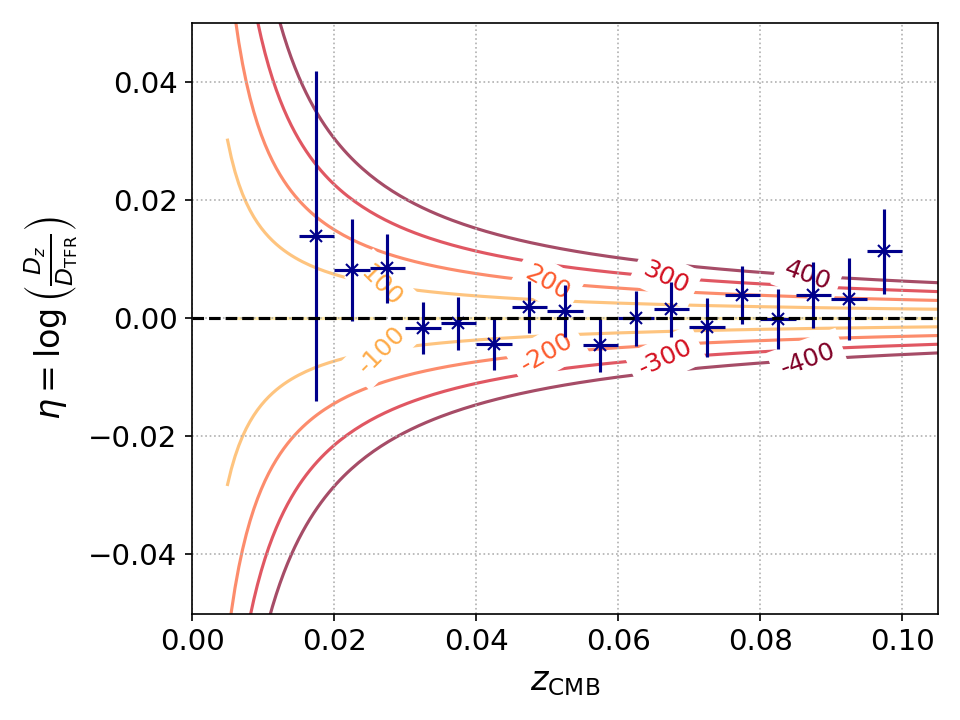}
  \caption{The weighted average log-distance ratio versus redshift for the DESI 
  DR1 PV Survey TF sample.  Lines of constant peculiar velocity are shown in 
  shades of orange/red.}
  \label{fig:PV_v_z}
\end{figure}

\begin{figure}
  \centering
  \includegraphics[width=0.49\textwidth]{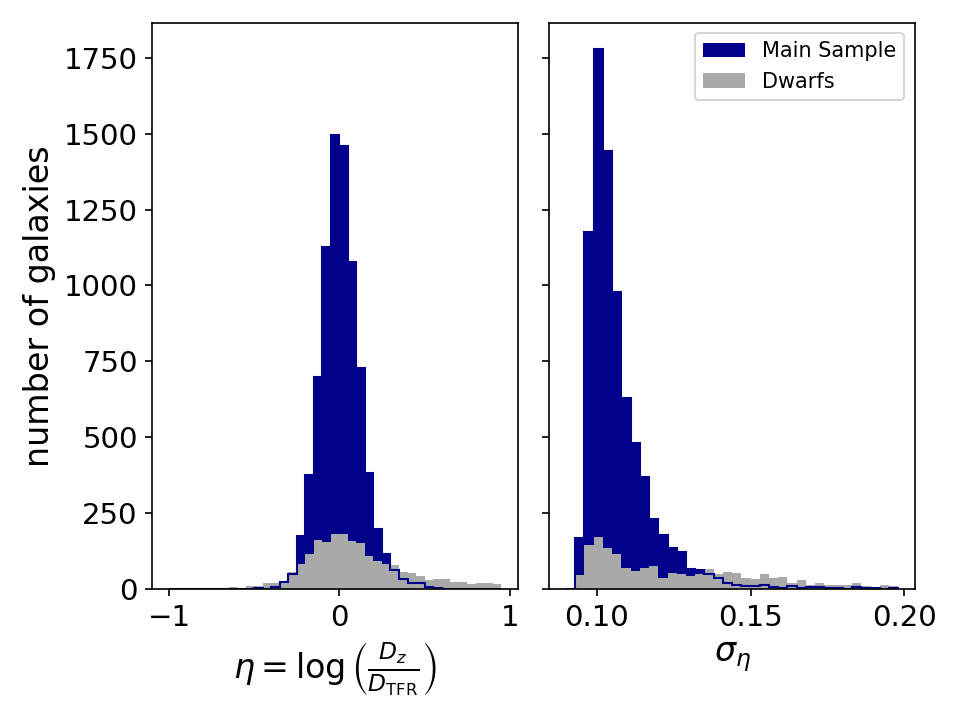}
  \caption{Distributions of the log-distance ratio (left) and uncertainty in the 
  log-distance ratio (right) for the DESI DR1 PV Survey TF sample.  Galaxies 
  shown in gray are those which are not part of the main cosmological sample as 
  defined in Sec.~\ref{sec:Alex_cuts}.}
  \label{fig:logdist_hists}
\end{figure}

The peculiar velocity of a galaxy alters its observed redshift, so the 
difference between the inferred and predicted galaxy distance moduli can be 
attributed to the peculiar velocity, along with some scatter that averages out 
over a large enough sample.  We use the calibrated TFR to estimate the distance 
modulus, $\mu_{\rm TF}$, given a galaxy's observed rotational velocity.  We 
convert the calibrated $y$-intercepts of the TFR in each redshift bin to an 
absolute magnitude using the central redshift of the bin, $z_{c,i}$, so that
\begin{equation}
  B_i = b_i - \mu(z_{c,i})
\end{equation}
A galaxy's predicted absolute magnitude is then calculated from the calibrated 
TFR as
\begin{equation}
  M_r = a\log \left( \frac{V(0.4R_{26})}{V_0} \right) + B_i
\end{equation}
using the $B_i$ of the redshift bin within which the galaxy falls.  We use the 
$B_i$ of the lowest redshift bin for those galaxies with $z < 0.03$, and we use 
the $B_i$ of the highest redshift bin for those galaxies with $z > 0.1$.  The 
TFR for the \Ntot galaxies in the DESI DR1 TF sample is shown in 
Fig.~\ref{fig:TFR}, where the absolute magnitudes for each object are calculated 
using the proper distance for a flat $\Lambda$CDM cosmology with 
$\Omega_M = 0.3151$, $H_0 = 100h$~km/s/Mpc using the DESI DR1 redshift of the 
center of the galaxy.

The effect of peculiar velocities directly on the observed magnitude of the 
galaxy is subdominant to the effect on redshift and intrinsic scatter of the 
TFR.  We extract the PV information by comparing the inferred distance modulus, 
$\mu (z_{\rm CMB})$, with that predicted by the TFR, $\mu_{\rm TF}$, where
\begin{equation}
  \mu_{\rm TF} = m_{r, {\rm corr}} - M_{r,{\rm TF}},
\end{equation}
to compute the log distance ratio, 
\begin{equation}
  \eta = 0.2\left(\mu (z_{\rm CMB}) - \mu_{\rm TF}\right).
\end{equation}
We choose to quantify the peculiar velocities with the log distance ratio 
because its errors are Gaussian-distributed.  In Fig.~\ref{fig:PV_v_z}, we show 
the weighted mean log distance ratio as a function of the redshift for the DESI 
DR1 TF sample.  The distance moduli and log distance ratios for the DESI DR1 TF 
sample are shown in Table~\ref{tab:pv}, and the distributions of the log 
distance ratios and their uncertainties are shown in 
Fig.~\ref{fig:logdist_hists}.

\begin{deluxetable}{CCCCCCCCCc}
\tablewidth{0pt}
\tablehead{\colhead{SGA-2020} & \colhead{R.A.} & \colhead{Decl.} & \colhead{Redshift} & \colhead{$D(26)$} & \colhead{$m_r(26)$} & \colhead{$V(0.4R_{26})$} & \colhead{$\mu$} & \colhead{$\eta$} & \colhead{Main} \\[-0.5em] \colhead{ID} & \colhead{[deg]} & \colhead{[deg]} & & \colhead{[arcmin]} & \colhead{[AB mag]} & \colhead{[km/s]} & \colhead{[AB mag]} & & \colhead{sample}}
\tablecaption{DESI DR1 TF catalog\label{tab:pv}}
\startdata
38   & 211.6620 & \phantom{-}39.5272 & 0.08099(3) & 0.42 & 17.50\pm0.03 & 122\phantom{.0}\pm11\phantom{.0} & 36.72\pm0.54 & \phantom{-}0.07\pm0.11 & True \\
55   & 199.7306 & \phantom{-0}1.7263 & 0.04400(3) & 0.50 & 17.58\pm0.02 & \phantom{0}81.5\pm\phantom{0}9.0 & 35.60\pm0.57 & \phantom{-}0.03\pm0.11 & True \\
117  & 138.9354 & \phantom{-0}2.6948 & 0.09004(4) & 0.57 & 16.68\pm0.03 & 181\phantom{.0}\pm14\phantom{.0} & 37.08\pm0.52 & \phantom{-}0.05\pm0.10 & True \\
223  & 197.7870 & \phantom{-}34.1713 & 0.03403(3) & 0.58 & 17.34\pm0.02 & \phantom{0}70\phantom{.0}\pm12\phantom{.0} & 34.82\pm0.72 & \phantom{-}0.07\pm0.14 & True \\
581  & 118.8598 & \phantom{-}18.9755 & 0.04332(3) & 0.95 & 16.19\pm0.02 & 134\phantom{.0}\pm11\phantom{.0} & 35.46\pm0.53 & \phantom{-}0.04\pm0.11 & True \\
669  & 198.7851 & \phantom{-0}3.2186 & 0.08359(3) & 0.62 & 15.86\pm0.01 & 293\phantom{.0}\pm17\phantom{.0} & 38.14\pm0.48 & -0.20\pm0.10 & True \\
731  & 334.3157 & \phantom{0}-0.0666 & 0.07714(3) & 0.56 & 17.55\pm0.02 & \phantom{0}82\phantom{.0}\pm12\phantom{.0} & 35.42\pm0.65 & \phantom{-}0.30\pm0.13 & True \\
1135 & 125.6008 & \phantom{-0}7.0371 & 0.07889(3) & 0.50 & 16.20\pm0.02 & 218.7\pm\phantom{0}7.0 & 37.40\pm0.46 & -0.08\pm0.09 & True \\
1158 & 162.9366 & \phantom{-0}3.0996 & 0.04644(2) & 0.78 & 15.61\pm0.02 & 178\phantom{.0}\pm11\phantom{.0} & 35.95\pm0.49 & -0.02\pm0.10 & True \\
1287 & 209.9386 & \phantom{0}-3.0886 & 0.02632(3) & 0.71 & 16.12\pm0.01 & 109\phantom{.0}\pm20\phantom{.0} & 35.09\pm0.74 & -0.10\pm0.15 & True
\enddata
\tablecomments{{Ten} of the \Ntot galaxies in the DESI DR1 TF catalog.  Sky positions and diameters of the 26 mag arcsec$^{-2}$ $r$-band isophote are from the SGA-2020 \citep{SGA}.  Redshifts are measured from the DESI DR1 spectra, and rotational velocities at $0.4R_{26}$ are computed as described in Sec.~\ref{sec:measure_rot_vel}.  Distance moduli are calculated from the calibrated TFR, and the log distance ratios are based on the difference between the observed and predicted distance moduli.  Table~\ref{tab:pv} is published in its entirety online in a machine-readable format.  A portion is shown here for guidance regarding its form and content.}
\end{deluxetable}


\subsection{Defining the main TF sample}\label{sec:Alex_cuts}

\begin{figure*}
  \centering
  \includegraphics[width=\textwidth, trim={0 0.5cm 0 0.5cm}, clip]{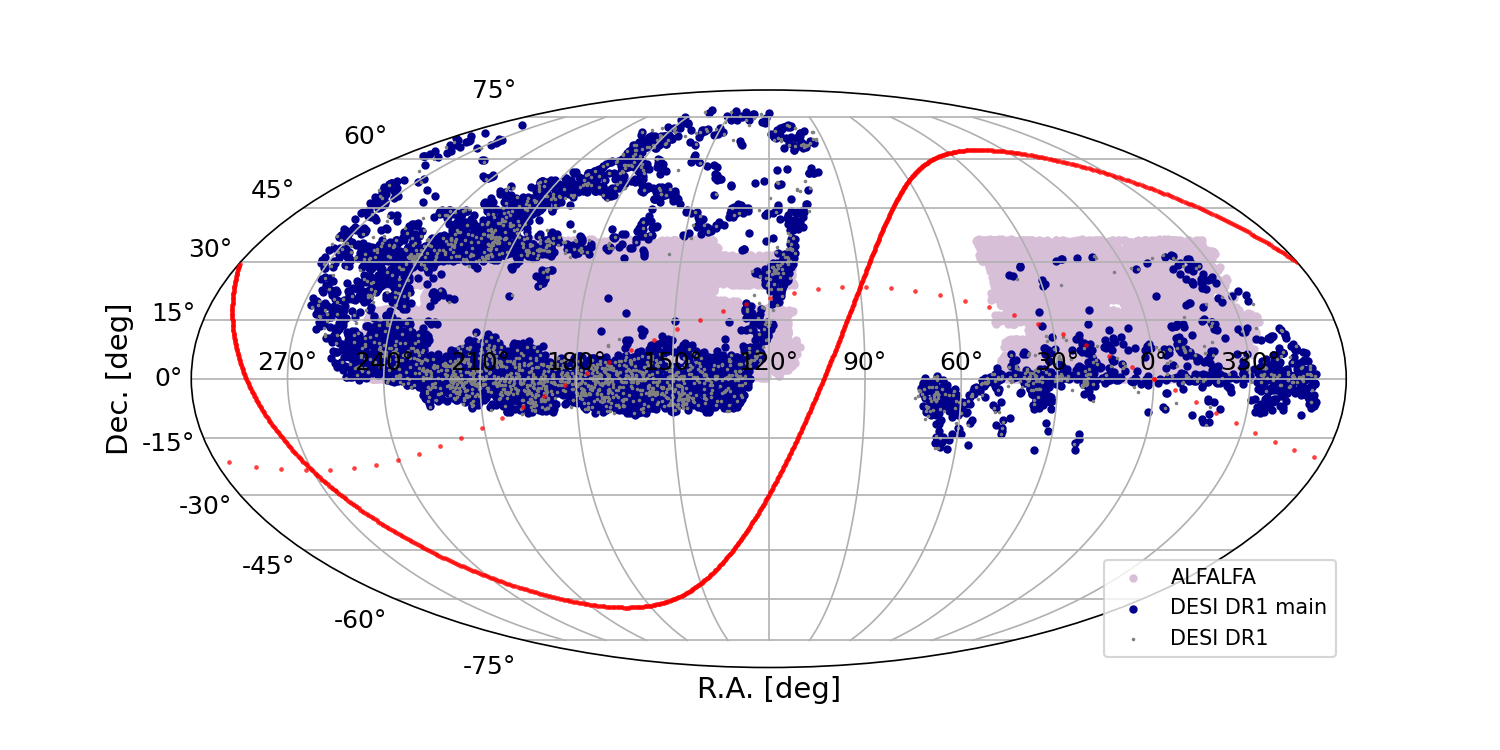}
  \caption{The distribution of Tully Fisher galaxies within the DESI DR1 
  dataset, presented here in a Mollweide projection.  The purple shaded region 
  corresponds to the ALFALFA \citep{Haynes2018} footprint, and the red solid 
  line represents the plane of the Milky Way.  The main DESI DR1 TF sample is 
  shown in dark blue, with those that do not satisfy the more stringent velocity 
  and/or magnitude requirements shown in grey.}
  \label{fig:sky_plot}
\end{figure*}

\begin{figure}
  \centering
  \includegraphics[width=0.5\textwidth]{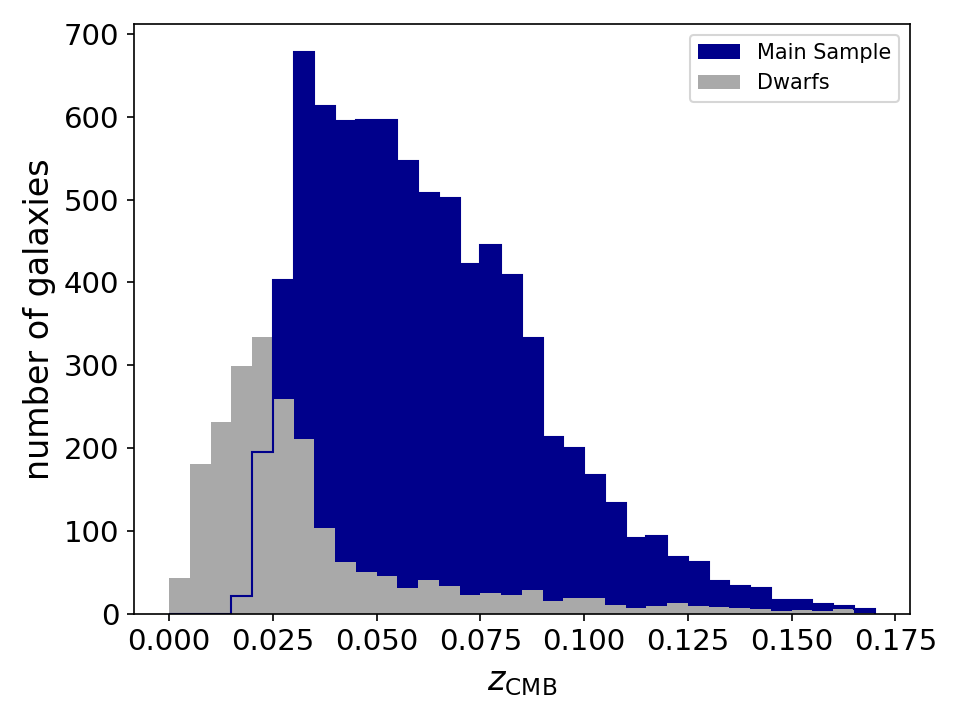}
  \caption{Redshift distribution of galaxies in the DESI DR1 PV Survey TF 
  sample.  The main DESI DR1 TF sample is shown in dark blue, with those that do 
  not satisfy the more stringent velocity and/or magnitude requirements shown in 
  gray.}
  \label{fig:z_dist}
\end{figure}

As is apparent in Fig.~\ref{fig:TFR}, the DESI DR1 TF sample has significant 
scatter, particularly above and to the left of the TFR.  These correspond to 
lower rotational velocities measured than what one would expect from the 
observed redshift and apparent magnitude.  One possibility for their 
underestimated rotational velocities is if their rotation curves have not 
reached the maximum velocity by $0.4R_{26}$.  As these objects will bias some 
studies using this sample, we extend a subset of our requirements for the 
calibration sample to the full TF sample, defining the main cosmology sample as 
those with
\begin{itemize}
  \item A rotational velocity $70\text{ km/s} < V(0.4R_{26}) < 300\text{ km/s}$ 
        and $\log V(0.4R_{26}) < 0.3(\mu(z_{\rm CMB}) - 34 - 5\log h) + 2$, and
  \item An apparent magnitude $m_r$ larger than the minimum of 
        $(17.75, \mu(z_{\rm CMB}) - 17 + 5\log h)$ (i.e., non-dwarf galaxies).
\end{itemize}
Galaxies in the DESI DR1 TF sample which are part of this main population are 
shown in the colored density bins in Fig.~\ref{fig:TFR}, while those that are 
not are shown as grey points.  The distribution of the TF galaxies in DESI DR1 
on the sky is shown in Fig.~\ref{fig:sky_plot}, and the $N(z)$ distribution of 
the DESI DR1 TF sample is shown in Fig.~\ref{fig:z_dist}.

\subsection{Constructing the clustering catalogs}

The positions and velocities of the DESI DR1 TF galaxy sample are used in 
subsequent analysis to constrain the growth rate of structure 
\citep{DESI_DR1_fs8_cf, DESI_DR1_fs8_ps, DESI_DR1_fs8_ml}.  For the purposes of 
these clustering measurements, we define a subsample of the DR1 TF catalog that 
is restricted to the most reliable peculiar velocity measurements, and those 
where we have better knowledge of the systematic impact of the DESI 
fiber-assignment process, density variations in the imaging used for target 
selection, and redshift failures.

We do this by first restricting to galaxies in the main TF sample (i.e., those 
shown in the colored density bins in Figs.~\ref{fig:TFR}--\ref{fig:PV_v_z} and 
described in Sec.~\ref{sec:Alex_cuts}).  We then cross-match these TF galaxies 
to a $z < 0.1$ sample of the BGS large-scale structure (LSS) catalogs presented 
in \cite{DESI_LSS} using each galaxy's \texttt{TARGETID}.  This restricts the TF 
data to only those objects within the well-defined angular mask of the LSS 
catalogs and allows us to borrow from those catalogs systematic weights 
accounting for completeness, imaging density fluctuations, and redshift 
success.  In doing this cross-match, we identify 31 galaxies where the redshifts 
reported for the galaxy in the BGS LSS catalogs differ by more than $10^{-4}$ 
from that given in the TF sample --- these arise because the redshift reported 
in the TF catalog is an average over all measurements taken at the center of 
that galaxy in both bright and dark conditions, while those in the BGS LSS 
catalog use only bright time measurements.  We elect to remove these galaxies 
from the clustering catalog rather than adopt either choice of redshift.

The end result is a subset of 6,807 galaxies (out of the full \Ntot) that we use 
for subsequent clustering measurements.  For this subset, we also produce random 
unclustered catalogs, with 20, 200, and $400\times$ the number of real galaxies 
with the same angular, redshift, and peculiar velocity error distributions.  
These are used to calculate and report the number density of BGS redshifts and 
TF-based peculiar velocity measurements at each TF galaxy's position in the 
clustering catalog \citep[used e.g., for measurements of the density and 
velocity power spectra in][]{DESI_DR1_fs8_ps}.  The clustering catalog and 
associated random catalogs are released along with the full TF sample.

\section{Discussion}\label{sec:discussion}

\subsection{Comparison with CosmicFlows-4}

\begin{figure}
  \centering
  \includegraphics[width=0.5\textwidth]{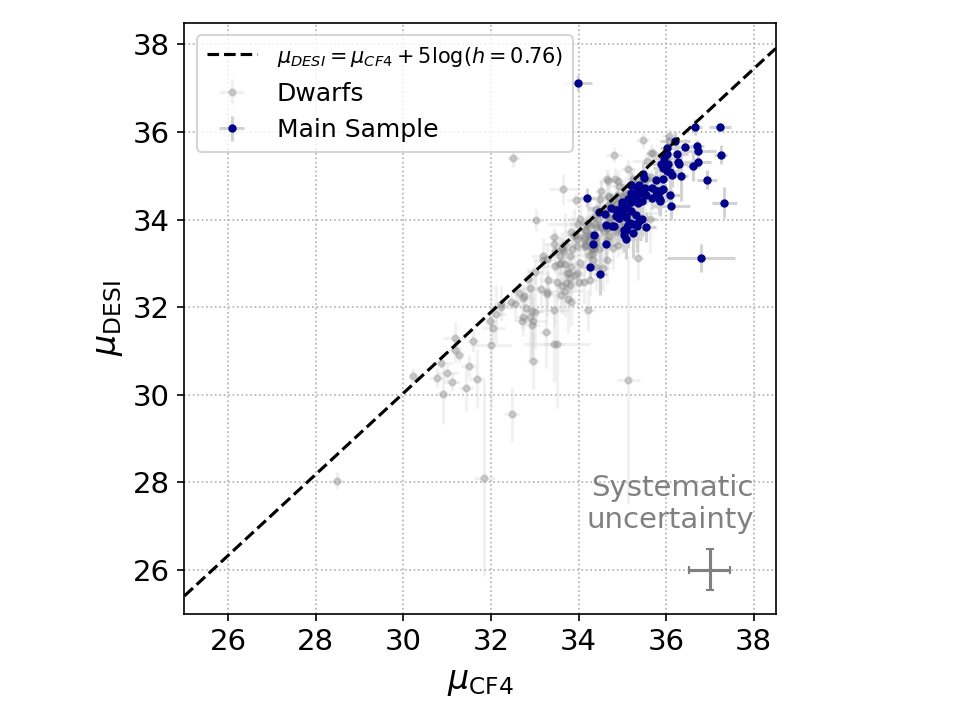}
  \caption{A comparison between the distance moduli from our calibrated TFR and 
  that of Cosmicflows-4 \citep[CF4;][]{Kourkchi2020a}.  Galaxies that are not 
  part of our main sample (defined in Sec.~\ref{sec:Alex_cuts}) are shown in 
  gray.  $y = x$ is shown in the black dashed line, adjusted for the value of 
  $H_0$ used in both catalogs.  We find relatively good agreement with CF4, with 
  a slight bias towards larger distance moduli ($0.21\pm 0.71$).}
  \label{fig:TFR_comparison}
\end{figure}

We can compare our TFR calibration with the Cosmicflows-4 TFR 
\citep[CF4;][]{Kourkchi2020a} for those objects in our sample that have velocity 
widths measured in the ALFALFA~\ion{H}{1} survey \citep{Haynes2018}.  While each 
of these catalogs contains $\sim$10,000 galaxies, ours covers a much smaller 
portion of the sky and extends to much higher redshifts (3580~sq.~deg., 
$z \lesssim 0.1$) than CF4 (almost full-sky, $z < 0.05$).  We show how our 
distance moduli compare with those assigned distance moduli in 
\cite{Kourkchi2020b} in Fig.~\ref{fig:TFR_comparison} for the 299 galaxies which 
are in both samples (96 of which are in our main cosmology sample), and we 
include the cross-matched sample in Table~\ref{tab:CF4_comp}.  The error bars 
shown in the figure, both the statistical (object-dependent) and systematic, 
reveal that our uncertainties are comparable to those of CF4.  After accounting 
for the different $H_0$ values used in the two catalogs, we find relatively good 
agreement with the CF4 TFR.  We note a slight bias of $0.21\pm 0.71$ towards our 
TFR predicting larger distance moduli, which we attribute to an offset in the 
zero-pointing of the two TFRs.  Of the three extreme outliers seen in 
Fig.~\ref{fig:TFR_comparison}, we find one to have a significantly different 
redshift ($\Delta z > 0.05$) between CF4 and DESI; the other two appear to have 
much larger \ion{H}{1} velocities relative to $V(0.4R_{26})$, making them 
outliers in the \ion{H}{1} TFR.  Visual inspection of their Legacy Surveys 
images and DESI DR1 spectra reveal no anomalies.

\begin{deluxetable}{CCCCCCCCCCCc}
\tablewidth{0pt}
\tablehead{\colhead{SGA-2020} & \colhead{R.A.} & \colhead{Decl.} & \colhead{Redshift} & \colhead{$D(26)$} & \colhead{$m_r$ (DESI)} & \colhead{$m_r$ (CF4)} & \colhead{$V(0.4R_{26})$ (DESI)} & \colhead{$Wmx$ (CF4)} & \colhead{$\mu$ (DESI)} & \colhead{$\mu$ (CF4)} & \colhead{Main} \\[-0.5em] \colhead{ID} & \colhead{[deg]} & \colhead{[deg]} &  & \colhead{[arcmin]} & \colhead{[AB mag]} & \colhead{[AB mag]} & \colhead{[km/s]} & \colhead{[km/s]} & \colhead{[AB mag]} & \colhead{[AB mag]} & \colhead{sample}}
\tablecaption{DESI DR1 -- CF4 cross-matched catalog\label{tab:CF4_comp}}
\startdata
2733  & 224.5346 & \phantom{-}3.4236 & 0.02150(2) & 1.42 & 14.91\pm0.02 & 14.7 & 119.9\pm\phantom{0}9.8           & 257\pm\phantom{0}5 & 33.84\pm0.55 & 34.39\pm0.48 & False \\
15509 & 141.0202 & \phantom{-}2.1216 & 0.03687(2) & 1.31 & 14.76\pm0.01 & 14.7 & 191.3\pm\phantom{0}7.5           & 415\pm29           & 35.32\pm0.49 & 36.73\pm0.54 & True \\
18021 & 211.6584 & \phantom{-}3.7487 & 0.04059(3) & 1.75 & 14.39\pm0.01 & 14.5 & 212\phantom{.0}\pm13\phantom{.0} & 425\pm12           & 35.31\pm0.52 & 36.28\pm0.49 & True \\
23113 & 329.6040 &           -0.7397 & 0.01627(8) & 1.40 & 15.00\pm0.00 & 14.6 & 101\phantom{.0}\pm25\phantom{.0} & 179\pm18           & 33.64\pm0.97 & 34.26\pm0.61 & False \\
24070 & 347.3241 & \phantom{-}0.0643 & 0.01574(2) & 1.28 & 15.19\pm0.03 & 15.2 & 107\phantom{.0}\pm13\phantom{.0} & 270\pm\phantom{0}8 & 33.64\pm0.62 & 34.29\pm0.48 & False
\enddata
\tablecomments{{Five} of the 299 galaxies that contain data from both DESI DR1 and Cosmicflows4.  Sky positions and diameters of the 26 mag arcsec$^{-2}$ $r$-band isophote are from the SGA-2020 \citep{SGA}.  Redshifts are measured from the DESI DR1 spectra, rotational velocities at $0.4R_{26}$ are computed as described in Sec.~\ref{sec:measure_rot_vel}, and distance moduli are calculated from the calibrated TFR.  The CF4 values are from Tables~1 and 4 of \cite{Kourkchi2020b}. Table~\ref{tab:CF4_comp} is published in its entirety online in a machine-readable format.  A portion is shown here for guidance regarding its form and content.}
\end{deluxetable}

\subsection{Anchoring the zero-point of the TFR}

The final step of calibrating the TFR is tying the zero-points in each redshift 
bin of the TFR to external calibrators with known independent distances.  The 
zero-points $B_i$ are unanchored and set such that the redshift-binned average 
distance modulus is 0 under the assumption that there is no average radial 
peculiar velocity over large sky areas.  This global zero-pointing results in an 
offset
\begin{equation}
  b_{\rm 0pt} = \Delta \mu_{\rm 0pt} = \langle \mu \rangle_{\rm TF} - \langle \mu \rangle_{\rm ext}
\end{equation}
to the TF Hubble diagram.  A joint fit of the DESI DR1 TF, FP, and supernova Ia 
distances from the SH0ES/Pantheon+ sample \citep{SH0ES22, Pplus_cosmo} is 
performed to find the global zero-point.  To improve the statistical uncertainty 
of the zero-point, we take advantage of galaxy groups to increase the overlap of 
supernova distance measurements with DESI DR1 PV Survey galaxies.  The method of 
zero-pointing the TF and FP samples is described in detail in 
\cite{DESI_DR1_zp}, including a description of the galaxy group catalog and 
$H_0$ constraints that follow from the zero-pointed distance-redshift relation.

\section{Conclusion}\label{sec:conclusion}

We calibrate the Tully-Fisher relation (TFR) using galaxies observed in the 
first data release (DR1) of DESI's eight year survey, targeted as part of DESI's 
Peculiar Velocity Survey.  This secondary targeting program measures the 
rotational velocities of \Ntot galaxies at $0.4R_{26}$ by placing fibers on the 
centers and major axes of spatially-resolved galaxies (primarily spirals) within 
the DESI footprint.

We fit for the TFR using \NbinGals galaxies across \Nbins redshift bins of width 
$\Delta z = 0.005$ from $0.03 < z < 0.10$ with 16 free parameters: the slope, 
$y$-intercepts of each of the \Nbins bins, and the intrinsic scatter in the TFR.  
We find a slope of \TFslope in the $r$-band for velocities at $0.4R_{26}$, and 
an intrinsic scatter of the TFR of \TFscatter.

Our calibrated TFR allows us to measure the peculiar velocity of \Ntot galaxies 
observed as part of the DESI PV Survey DR1 TF sample.  While the target 
selection used to define this sample of \Ntot galaxies was designed to identify 
late-type spiral galaxies, we find some interloping ellipticals, lenticulars, 
and irregular galaxies.  We provide morphological classifications from a 
self-supervised learning algorithm for the \Ntot galaxies for further sample 
refinement.  Along with the Fundamental Plane sample of the DESI PV Survey DR1 
\citep{DESI_DR1_fp}, we use these peculiar velocities to study the growth rate 
of structure \citep{DESI_DR1_fs8_cf, DESI_DR1_fs8_ps, DESI_DR1_fs8_ml} and 
estimate the value of the Hubble constant, $H_0$ \citep{DESI_DR1_zp}.  This TF 
catalog comprises $\sim$20\% of the full catalog expected by the completion of 
the full 8-year DESI PV Survey, which will comprise the largest catalog of TF 
distances to date.  The substantial increase in sample size will greatly reduce 
the statistical uncertainties in the measured PVs of the sample, significantly 
contributing to the overall increase in precision of the growth rate of 
structure forecast for the DESI PV Survey \citep{Saulder2023} in the local 
universe.  We will also be able to study the systematics of the TFR in 
unprecedented detail with these future samples, increasing our understanding of 
the TFR and its use in measuring distances.

\section*{Data Availability}

Data shown in figures, as well as example Python code to generate the figures, 
will be available on Zenodo upon acceptance.

\begin{acknowledgements}

The authors thank former undergraduate students Hayley Nofi (Villanova 
University), MJ Keller (University of Rochester), and Sunny Woo (University of 
Rochester) for their help in various stages of this analysis.  This material is 
based on work supported in part by the National Science Foundation Grant No. 
PHY-2149332.

This material is based upon work supported by the U.S. Department of Energy 
(DOE), Office of Science, Office of High-Energy Physics, under Contract No. 
DE–AC02–05CH11231, and by the National Energy Research Scientific Computing 
Center, a DOE Office of Science User Facility under the same contract.  
Additional support for DESI was provided by the U.S. National Science Foundation 
(NSF), Division of Astronomical Sciences under Contract No. AST-0950945 to the 
NSF’s National Optical-Infrared Astronomy Research Laboratory; the Science and 
Technology Facilities Council of the United Kingdom; the Gordon and Betty Moore 
Foundation; the Heising-Simons Foundation; the French Alternative Energies and 
Atomic Energy Commission (CEA); the National Council of Humanities, Science and 
Technology of Mexico (CONAHCYT); the Ministry of Science, Innovation and 
Universities of Spain (MICIU/AEI/10.13039/501100011033), and by the DESI Member 
Institutions: \url{https://www.desi.lbl.gov/collaborating-institutions}.  Any 
opinions, findings, and conclusions or recommendations expressed in this 
material are those of the author(s) and do not necessarily reflect the views of 
the U. S. National Science Foundation, the U. S. Department of Energy, or any of 
the listed funding agencies.

The authors are honored to be permitted to conduct scientific research on 
I'oligam Du'ag (Kitt Peak), a mountain with particular significance to the 
Tohono O’odham Nation.

\end{acknowledgements}

\software{
Astropy \citep{astropy:2013, astropy:2018, astropy:2022}, 
Corner \citep{corner}, 
HyperFit \citep{Robotham2015}, 
Kcorrect \citep{Blanton2007},
Matplotlib \citep{matplotlib},
NumPy \citep{numpy}, 
SciPy \citep{scipy}
}

\bibliographystyle{aasjournalv7}
\bibliography{Doug1225_sources}

\end{document}